# Fate of the scalar quartic couplings in the inert models

Priyotosh Bandyopadhyay [*] and Pram Milan P. Robin [†]
*Indian Institute of Technology Hyderabad, Kandi, Sangareddy-502285, Telangana, India*

In this article we consider three inert models, namely the Inert Singlet Model (ISM), Inert Triplet Model (ITM), and Inert Doublet Model (IDM) as beyond Standard Model scenarios, and look into the running of the scalar quartic couplings at one- and two-loop levels. Interestingly, the Landau poles at one-loop and Fixed Points at two-loop have been observed for these scenarios, very similar to the $\phi^4$ theory, which is absent in the Standard Model. The role of Higgs portal couplings in attaining these features has been examined. For inert singlet and triplet models, the portal couplings give rise to terms without any residual phases, enhancing this Fixed Point behavior. In the case of the Inert Doublet Model, $\lambda_{4,5}$ terms which have the residual phases, spoil this behavior. Finally, we consider perturbative unitarity constraints to put limits on the scalar quartic couplings for various perturbativity scales in both one- and two-loop. Larger portal couplings, i.e., $\gtrsim 2$, get strong perturbative bounds as low as $10^{2-3}$ GeV. Uncertainties of the Fixed Points, and phenomenological aspects of such scenarios are also briefed.

## I. INTRODUCTION

The Standard Model (SM) of particle physics declared its supremacy when the Higgs boson was discovered at the Large Hadron Collider in 2012 [1,2]. However, the SM stumbles upon various issues: as it does not contain a dark matter candidate, neutrinos are massless in SM which the oscillation data proves otherwise, SM electroweak vacuum is metastable [3–6] and a first-order phase transition (FOPT) [7–9] is not possible within SM, and so on. In this article, we consider the inert models with the scalar extensions which provide the much-needed dark matter, the vacuum stability, and also promise to have a possible FOPT. The article focuses on the other theoretical bounds coming from one- and two-loop $\beta$-functions addressing possible Landau poles (LPs), Fixed Points (FPs) and finally constraining the parameter space from perturbative unitarity.

In literature, the scalar extensions are studied concerning the dark matter with the simple extension of SM, viz., an Inert Singlet Model (ISM) [10–17], Inert Doublet Model (IDM) [18–37], and Inert Triplet Model (ITM) [16,25,26,38–41]. These models are also motivated to stabilize the electroweak vacuum [32,41–49] and the possibility of first-order phase transition is also looked into in [7–9,34,47,50–59]. However, theoretical bounds on the higher electroweak values of the quartic couplings can come from the perturbativity [21,22,32,44,46–48]. There are some studies on the perturbativity of these scalar extensions [21,22,32,46–48], especially the Landau poles in the scalar sector [21,35,48,60,61] and Fixed Points in the gauge sector [62]. However, a comparative study of different inert scalar extensions, which was missing, is the topic of this article. In particular, we show how the portal couplings are crucial for these behavior, and how the scalars from different gauge representations behave similarly or differently.

In this study, we first look into the occurrence of the Landau poles (LPs) and the Fixed Points (FPs) at one- and two-loop levels, respectively for the inert models. Then we use perturbative unitarity constraints to bound the scalar quartic couplings and corresponding perturbativity scales at one- and two-loop using the corresponding RG evolutions. Fixed Points in RG evolution play a crucial role in understanding the stability and behavior of quantum field theories at high energy scales. There are various theoretical reasons for studying FPs. The occurrence of FP suggests that the theory becomes scale-invariant at some energy scales. This can lead to relations between various couplings and it can help in reducing the number of parameters in the theory. References [63,64] show that FPs can give insights into underlying symmetries in the theory. This could also help to draw connections with higher symmetries like conformal symmetries. However, in this article, we restrict ourselves to a bottom-up approach where we are interested in the roles of different scalar quartic couplings for the inert extensions of SM in attaining the FPs at two-loop order and corresponding bounds of perturbativity.

[*]Contact author: bpriyo@phy.iith.ac.in
[†]Contact author: ph22resch01001@iith.ac.in

The perturbative calculation of $\phi^4$ theory up to seven-loop shows alternating appearance of Landau poles and Fixed Points for every odd- and even-loop of the beta-functions [65,66]. However, these ultraviolet (UV) Fixed Points are not reliable; in fact, in an $O(N)$ symmetric real scalar $\phi^4$ theory in four dimensions, there is no UV FP [67,68]. In this article, we restrict ourselves to studying the one- and two-loop beta-functions of the quartic couplings. Unlike the $\phi^4$ theory, the Standard Model does not have any LP at the one-loop and FP at the two-loop level at higher energy scales [69]. This is attributed to the fact that the SM parameters are fixed by the experimental values at the initial electroweak scale.

Nonetheless, in the extension of SM with inert scalars this behavior of LP and FP reappear at one- and two-loop, respectively. For ISM we introduce two new quartic couplings $\lambda_S$ (self-coupling) and $\lambda_{HS}$ (portal coupling). Likewise, in ITM, we get additional quartic couplings $\lambda_T$ (self-coupling of triplet $T$) and $\lambda_{HT}$ (portal coupling). These additional new parameters can be varied and it can lead to the possibility of FPs in these extensions of the SM, which can be ascribed to the potential without any residual phase. In the case of IDM, we have four more additional quartic couplings namely the self-coupling of the inert doublet $\lambda_2$ and three portal couplings $\lambda_3, \lambda_4$, and $\lambda_5$. The terms associated with $\lambda_{4,5}$ have residual phases and thus behave as a spoiler to achieve FP at two-loop, while $\lambda_3$ behaves similarly to the portal couplings in ISM and ITM. Finally, we constrain the parameter space in the plane of quartic couplings and the perturbativity scale at one- and two-loop by considering perturbative unitarity bounds [21,22,60,67,70–74].

In this paper, first, we provide an overview of the possibility of FPs in a simple scalar theory like $\phi^4$-theory in Sec. II. In Sec. III, we briefly discuss the absence of FPs in SM. Then we focus on inert scalar extensions in Sec. IV. We study the occurrence of FPs at two-loop and put the corresponding bounds from perturbativity at one and two-loop respectively. Similarly, in Sec. IV B we studied the Inert Triplet extension, and in Sec. IV C, the Inert Doublet Model is described where we emphasize the portal couplings which spoil the Fixed Point behavior. Uncertainties are briefed in Sec. V and phenomenological aspects has been summarized in Sec. VI. Finally, in Sec. VII we summarize and conclude.

## II. BEHAVIOR OF THE QUARTIC COUPLING IN $\phi^4$ THEORY

In quantum field theory, $\phi^4$-theory is the simplest example of an interacting scalar field theory. It is a prototype for studying renormalization group (RG) flows, where a Landau pole at one-loop and a Fixed Point at two-loop is observed. In this section, we briefly discuss the $\beta$—functions in $\phi^4$-theory. The tree-level Lagrangian for $\phi^4$-theory is given by,

$$\mathcal{L} = (\partial_\nu \phi)(\partial^\nu \phi) - \frac{m^2}{2}\phi^2 - \frac{\lambda}{4!}\phi^4, \tag{2.1}$$

where $\phi$ is a real scalar field. $m^2$ and $\lambda$ are the bare mass-squared parameter and quartic coupling respectively. For our analysis, we assume $m^2 > 0$ and $\lambda > 0$, at a reference energy scale $\mu_0$. As a result of the renormalization procedure, the quartic coupling $\lambda$ changes with the energy scale $\mu$, and that evolution is governed by the renormalization group (RG) equation, $\mu \frac{d\lambda(\mu)}{d\mu} = \beta(\lambda)$, where $\beta(\lambda)$ is the $\beta$-function for the coupling $\lambda$.

The $\beta$-function of $\lambda$ is defined as a perturbative series expansion in $\lambda$ in Eq. (2.2).

$$\beta(\lambda) = \mu \frac{d\lambda}{d\mu} = \lambda \sum_{k=1}^{\infty} c_k \lambda^k, \tag{2.2}$$

where $c_k$'s are determined from Feynman diagram calculations. The n-loop beta function, denoted as $\beta^{(n)}$ is given by setting the upper limit of the above series as $n$.

$$\beta^{(n)}(\lambda) = \lambda \sum_{k=1}^{n} c_k \lambda^k, \tag{2.3}$$

where the $n$-loop beta function is a polynomial of degree $n+1$. Figure 1 represents one- and two-loop diagrams contributing to the corresponding $\beta$-functions. The one-loop and two-loop $\beta$-functions are given by,

$$\beta^{\text{1-loop}}(\lambda) = \frac{3\lambda^2}{16\pi^2} \tag{2.4}$$

and

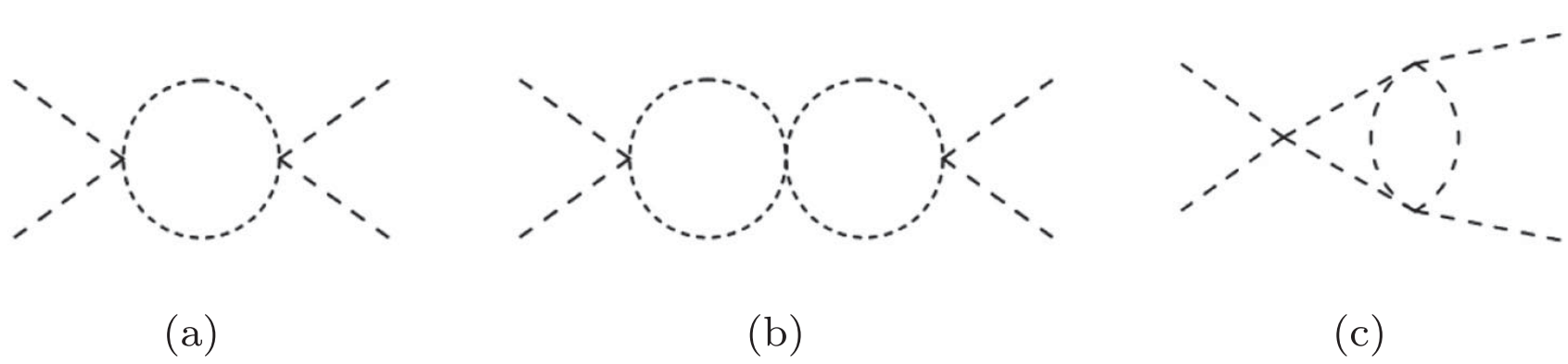

FIG. 1. Sample Feynman diagrams that contribute to the beta-functions in $\phi^4$ theory. (a) is at one-loop level, (b) and (c) are at two-loop level.

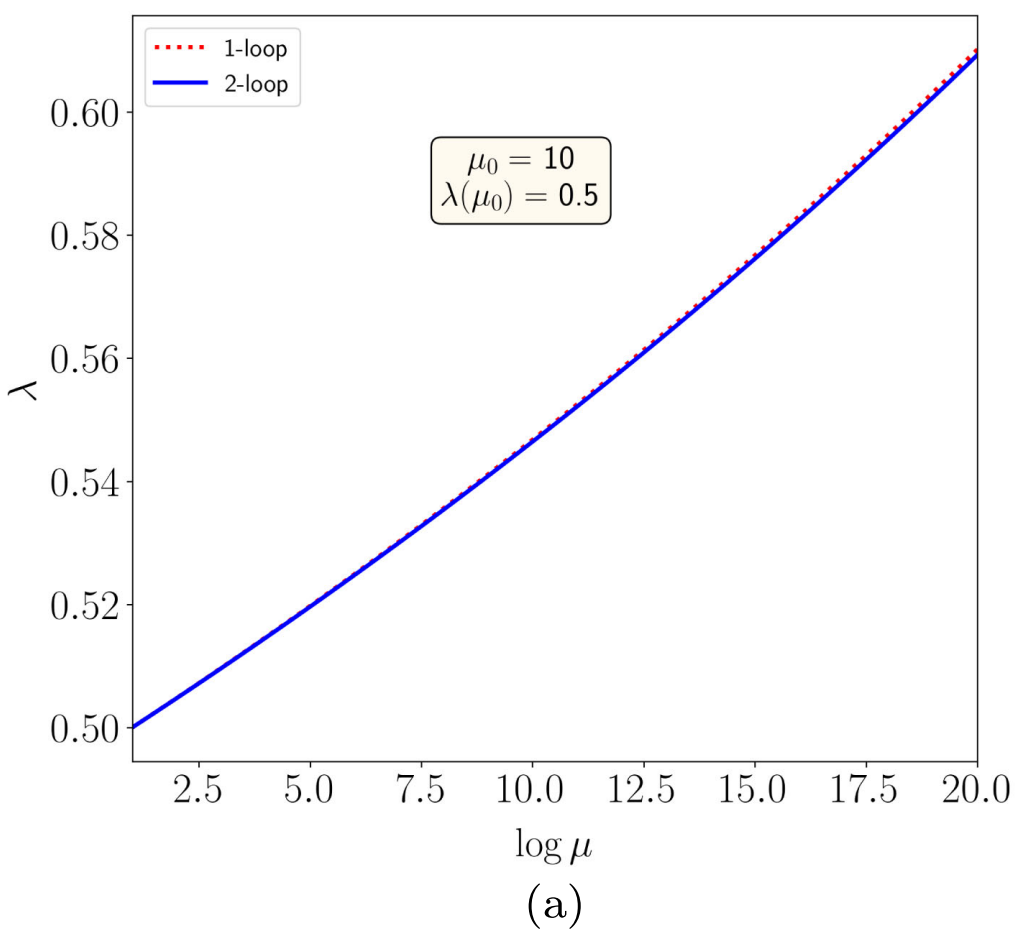

(a)

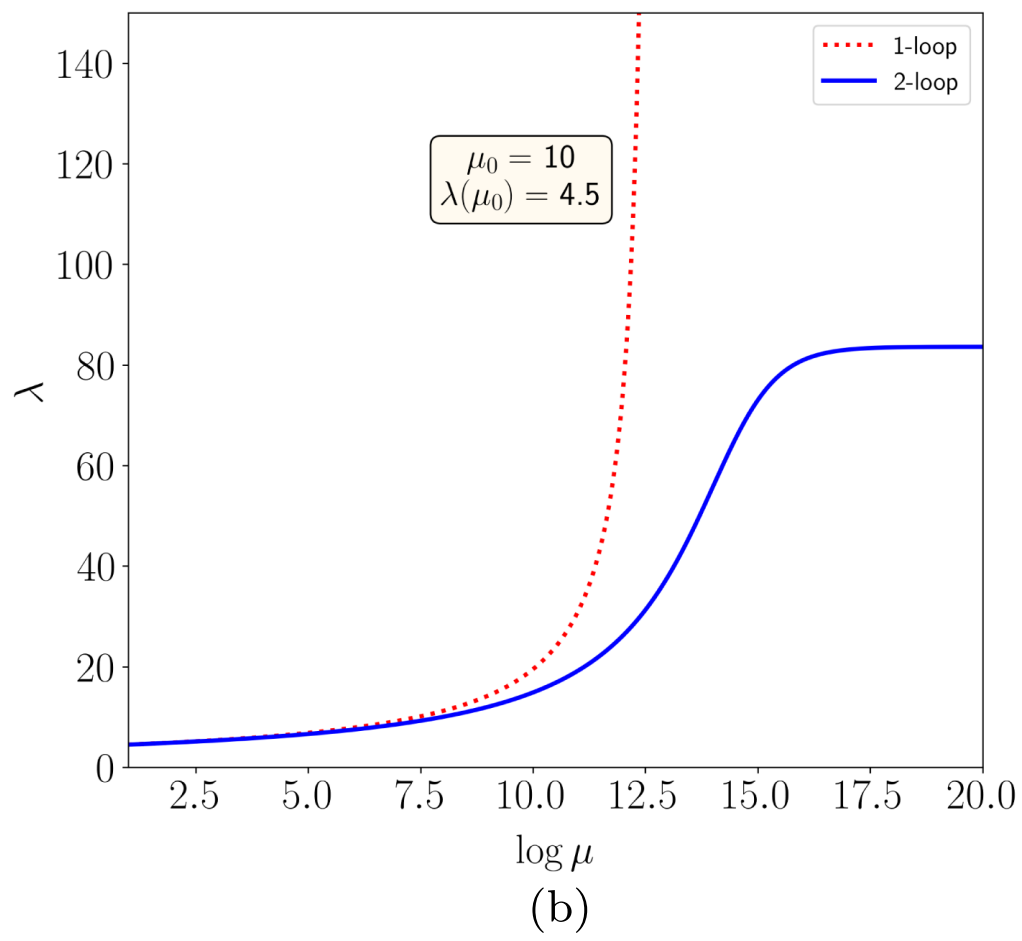

(b)

FIG. 2. RGE evolution of quartic coupling $\lambda$ in $\phi^4$-theory for one- and two-loop orders for (a) the initial value of $\lambda = 0.5$ at $\mu = 10$ GeV and (b) $\lambda = 0.5$ at $\mu = 4.5$ GeV, respectively.

$$\beta^{2\text{-loop}}(\lambda) = \frac{3\lambda^2}{16\pi^2} - \frac{17}{3}\frac{\lambda^3}{(16\pi^2)^2}. \tag{2.5}$$

The behavior of the $\beta$-functions often gives different theoretical limits. One such exciting behavior is the existence of Fixed Points (FPs). Fixed Points in RG evolution are regions where the $\beta$-function vanishes and the coupling does not change under scale transformations, indicating scale invariance. The FP is called a Gaussian FP, if $\lambda = 0$ (where the theory becomes a free theory); and non-Gaussian, if $\lambda \neq 0$ (corresponds to interacting theories that are scale-invariant). At the one-loop level, the zeros of the beta function can be obtained by setting $\beta^{1\text{-loop}}(\lambda) = 0$. This gives the Gaussian Fixed Point, $\lambda = 0$. At the one-loop level, there are no other Fixed Points. However, at the two-loop level, the zeros of the beta function are given by setting $\beta^{2\text{-loop}}(\lambda) = 0$.

$$\beta^{2\text{-loop}}(\lambda) = \frac{\lambda^2}{16\pi^2}\left[3 - \frac{17\lambda}{3(16\pi^2)}\right] = 0, \tag{2.6}$$

which gives,

$$\lambda = 0 \quad \text{and} \quad \lambda = \frac{9}{17}(16\pi^2) \approx 83.60. \tag{2.7}$$

Landau pole (LP) is another feature that can be present in an RG evolution. It is the energy scale at which the coupling strength of a quantum field theory becomes infinite. The Landau pole in the one-loop can be obtained by solving the one-loop RG equation. At an LP, the $\lambda(\mu)$ grows without bound at some finite energy scale as can be seen from Eq. (2.9).

$$\mu\frac{d\lambda}{d\mu} = b_1\lambda^2, \qquad \Rightarrow \int_{\lambda_0}^{\lambda}\frac{d\lambda'}{(\lambda')^2} = \frac{3}{16\pi^2}\int_{\mu_0}^{\mu}\frac{d(\mu')}{\mu'}, \tag{2.8}$$

and finally,

$$\lambda(\mu) = \frac{\lambda_0}{1 - \frac{3\lambda_0}{16\pi^2}\ln\left(\frac{\mu}{\mu_0}\right)}, \tag{2.9}$$

where $\mu_0$ is the reference scale and $\lambda_0 = \lambda(\mu_0)$. The denominator vanishes at a finite energy scale $\mu = \mu_0 \exp(\frac{16\pi^2}{3\lambda_0})$ and hence $\lambda(\mu)$ blows up. The RG evolution of $\lambda$ is shown in Fig. 2. In Fig. 2(a) for $\lambda = 0.5$ at $\mu_0 = 10$ GeV, we do not see any LP or FP at one- and two-loop levels. However, in Fig. 2(b) for $\lambda = 7$ at $\mu_0 = 10$ GeV we get an LP (shown in the dotted-red curve) at one-loop and an FP at two-loop (solid blue curve). It is evident that the two-loop Fixed Point is approaching $\lambda = 83.60$, which is independent of the initial values of $\lambda$ for those in which we observe an FP. However, the energy at which the Fixed Point behavior begins will decrease as the initial value of $\lambda$ increases. The detailed analysis of LPs and FPs for $\phi^4$ theory, till the seven-loop level has been carried out in [65,66,75] and shown that the FPs appearing in $\phi^4$-theory are not perturbatively reliable. The nonperturbative analysis of the absence of UV FPs in an $O(N)$ symmetric $\phi^4$ theory in four dimensions is described in [67,68]. In this article, we look into the behavior of scalar quartic coupling at one- and two-loop levels for the inert models, namely inert singlet, inert doublet, and inert triplet.

## III. ABSENCE OF LANDAU POLES AND FIXED POINTS IN THE STANDARD MODEL

It is obvious that the simplicity of $\phi^4$-theory cannot be expected in a highly nontrivial non-Abelian gauge theory like the Standard Model, especially where the theory is measured experimentally with high precision at the electroweak (EW) scale. The RG evolution of SM is well studied focusing on various aspects of the theory. It is worth mentioning that there are no FPs or LPs poles in the RG evolution of parameters in SM, especially the scalar quartic coupling $\lambda$. The one- and two-loop $\beta$-functions of $\lambda$ are functions of itself and other parameters like the Yukawa couplings and gauge couplings. The one-loop $\beta$-function of $\lambda$ is given in Eq. (3.1) and it can be seen that the maximum negative contribution comes from the Yukawa term as depicted in Fig. 3(a), which drives $\lambda$ toward the negative values. The corresponding two-loop $\beta$-functions are given in Appendix A 1

$$\beta_\lambda^{\rm SM} = \frac{1}{(4\pi)^2}\left(\frac{27}{800}g_1^4 + \frac{9}{80}g_1^2 g_2^2 + \frac{9}{32}g_2^4 - \frac{9}{20}g_1^2\lambda + 6\lambda^2 + 3\lambda Y_t^2 - \frac{9}{4}g_2^2\lambda - \frac{3}{2}Y_t^4\right) \tag{3.1}$$

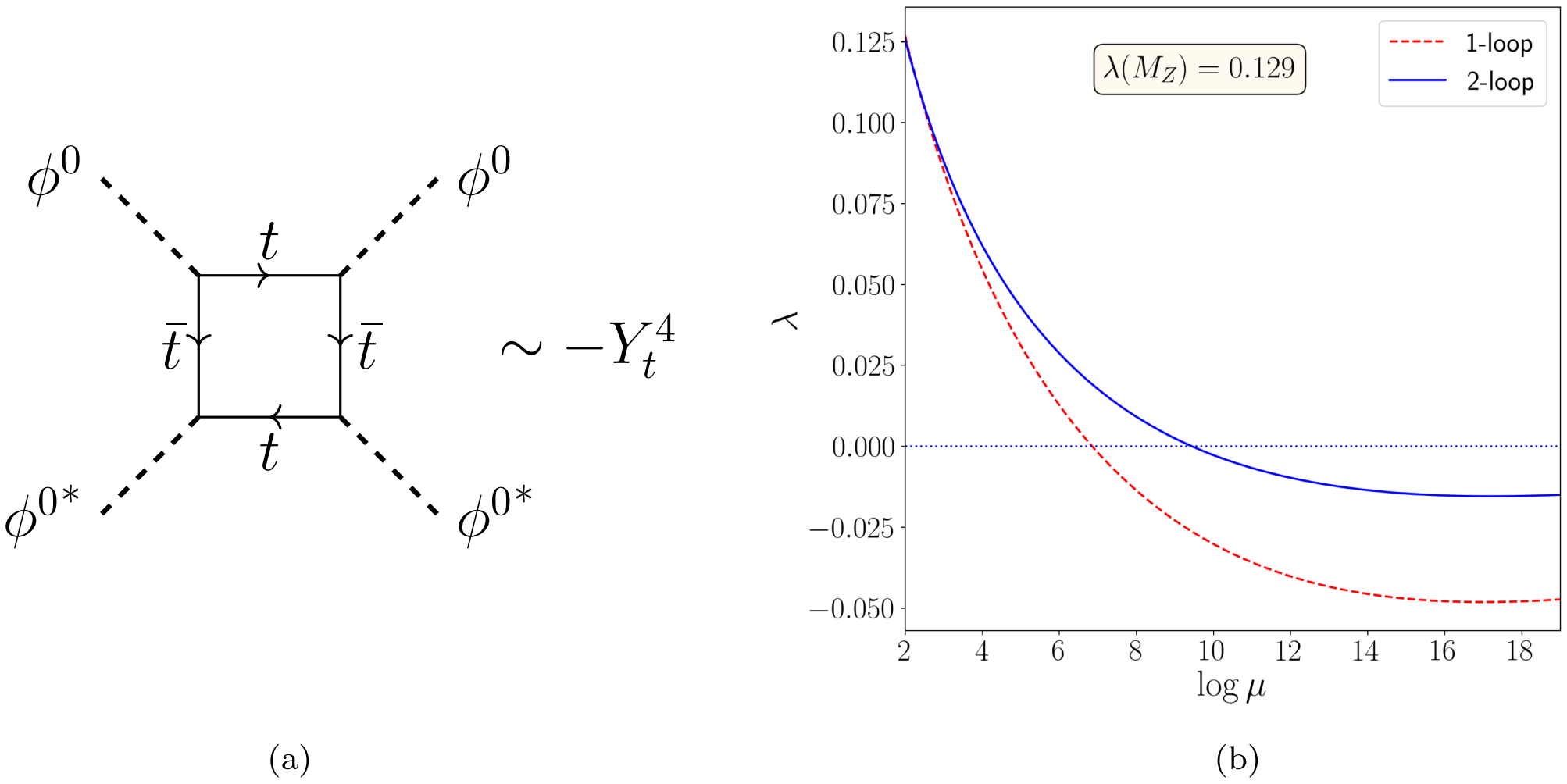

FIG. 3. (a) The top-loop diagram contributing negatively to the Higgs quartic coupling. (b) RG evolution of SM quartic coupling $\lambda$ verses scale $\mu$ at one- and two-loop order. The initial value of Higgs-quartic coupling $\lambda_H(M_Z) = 0.129$ at tree-level.

Even though these $\beta$-functions are also polynomials, we do not have any freedom in choosing the initial values of the parameters as they are experimentally measured with impeccable precision. With the standard values of the parameters at the electro-weak scale at one-(two-) loop, with $\lambda = 0.1271$ (0.12604), $g_1 = 0.4639$ (0.4626), $g_2 = 0.6476$ (0.6478), $g_3 = 1.1666$ (1.1666), $Y_t = 0.9495$ (0.9369) while other Yukawa couplings are neglected, the RG evolution of SM Higgs quartic coupling $\lambda$ is shown in Fig. 3(b). It can be seen that both in one- and two-loop evolutions we do not have any LP or FP, rather the running coupling $\lambda$ becomes negative at some high energy making the theory unbounded from below and the vacuum unstable. The addition of scalars from different gauge representations can bring back the stability of the electroweak vacuum [29,32,46,48].

However, when we extend the SM with new fields, additional parameters arise and we may be able to find LP and FP behavior for some parameter values. In the following sections, we examine a few inert scalar extensions that can exhibit the LP and FP behavior at one- and two-loop levels, at least for some regions in the parameter space. Along with this observation we also draw bounds from perturbative unitarity [60,67,70–74] at one- and two-loop levels for different initial values of new physics couplings. The evolution of couplings based on the perturbative series expansion of $\beta$-functions will have to satisfy perturbative unitarity at the corresponding energy scales given by

$$\lambda_i(\mu) < 4\pi, \qquad g_i < 4\pi, \qquad Y_i < \sqrt{4\pi}, \tag{3.2}$$

where $\lambda_i$ are the Higgs quartic couplings in a model, $g_i$ are the gauge couplings and $Y_i$ are the Yukawa couplings. Here we use the above criteria Eq. (3.2) obtained from the unitarity of scattering cross sections that are experimentally measurable [60,67,70–74]. For our studies in the inert models we considered all quartic couplings to satisfy $\lambda_i(\mu) < 4\pi$, all gauge couplings to satisfy $g_i < 4\pi$, and all Yukawa couplings to satisfy $Y_i < \sqrt{4\pi}$ for the perturbative unitarity.

## IV. INERT SCALAR EXTENSIONS OF THE STANDARD MODEL

In the following subsections, we study three inert scalar extensions of the Standard Model that give us a number of new parameters, namely the new mass and the quartic couplings. Additional fields and vertices can change the dynamics of the $\beta$-functions and hence affect the behavior of RG evolution of the couplings including the SM quartic coupling $\lambda$. This leads to the appearance of LPs and FPs at

one- and two-loop levels in some parameter regions. The interesting observation is that certain quartic couplings enhance the possibility whereas some others lessen those effects. In this article, we analyse such behavior for the inert models for three different representations of $SU(2)$, namely, Inert Singlet extension, Inert Doublet extension, and inert triplet extension. Moreover, here we are not talking about FPs in the most general sense like in a dynamical system where all the first-order dynamical equations vanish. We are only focussing on the scalar sector and particularly the evolution of the SM-like Higgs quartic coupling $\lambda$. Later we constrain the parameter space of scalar quartic couplings at one- and two-loop using perturbative unitarity [60,67,70–74] as defined in Eq. (3.2).

### A. Inert Singlet Model

The simplest scalar extension of the Standard Model (SM) is the Inert Singlet Model (ISM), where the singlet field $S$ does not take part in electroweak symmetry breaking. In this model, an additional complex scalar field $S$ is introduced along with the SM Higgs doublet $\Phi$. The new field $S$ is singlet under SM gauge groups. Additionally, it is odd under a global $\mathbb{Z}_2$ symmetry, while the SM particles are even. This $\mathbb{Z}_2$ odd nature forbids the $S$ to participate in the EWSB, which is the reason for the term "inert" and also makes it a viable dark matter candidate [10–15,17]. The scalar potential is given by,

$$V(\Phi, S) = \mu^2\Phi^\dagger\Phi + m_S^2 S^* S + \lambda_H(\Phi^\dagger\Phi)^2 + \lambda_S(S^*S)^2 + \lambda_{HS}\Phi^\dagger\Phi S^* S, \tag{4.1}$$

where $\Phi = (\phi^+ \phi^0)^T$, $S = \frac{1}{\sqrt{2}}(S_R + iS_I)$, $m_S$ is the singlet mass, and $\lambda_H$ is the SM-like Higgs coupling which is equivalent to $\lambda$ in the SM. $\lambda_S$ is the singlet self-coupling and $\lambda_{HS}$ is the Higgs-singlet portal coupling. As $S$ does not mix with the SM Higgs boson, it gives rise to a complex scalar mass eigenstate, which is also a candidate dark matter. The corresponding quartic couplings should satisfy the following stability conditions [76,77].

$$\lambda_H \geq 0, \qquad \lambda_S \geq 0, \qquad \lambda_{HS}^2 - 4\lambda_H\lambda_S > 0. \tag{4.2}$$

As we are interested in the behavior of the RG evolution of the quartic couplings, we study their scale dependency by calculating the corresponding $\beta$-functions through SARAH 4.15.1 [78]. The complete set of $\beta$-functions for ISM is given in the Appendix A 2.

We can see a positive contribution proportional to $\lambda_{HS}^2$ coming from the one-loop scalar diagram in Fig. 4 to the one-loop $\beta$-function given in Eq. (4.3). Some negative contributions can be found at two-loop $\beta$-function proportional to $\lambda_H\lambda_{HS}^2, \lambda_{HS}^3$ which are suppressed by a loop factor of $\frac{1}{(4\pi)^4}$.

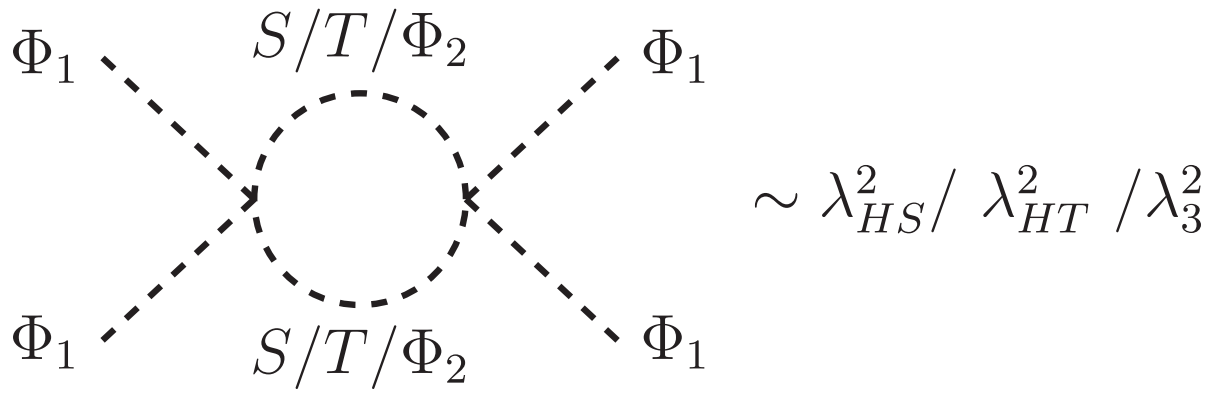

FIG. 4. One-loop Higgs-portal diagram in ISM.

$$\begin{aligned}
\beta_{\lambda_H}^{\mathrm{ISM\,(1\text{-}loop)}} &= \beta_{\lambda_H}^{\mathrm{SM\,(1\text{-}loop)}} + \frac{1}{(4\pi)^2}\left(\frac{1}{16}\lambda_{HS}^2\right)\\
\beta_{\lambda_H}^{\mathrm{ISM(2\text{-}loop)}} &= \beta_{\lambda_H}^{\mathrm{SM}} + \frac{1}{(4\pi)^2}\left(\frac{1}{16}\lambda_{HS}^2\right)\\
&\quad - \frac{1}{(4\pi)^4}\left(\frac{5}{128}\lambda_H\lambda_{HS}^2 + \frac{1}{64}\lambda_{HS}^3\right).
\end{aligned} \tag{4.3}$$

The evolution of SM-like Higgs quartic coupling $\lambda_H$ in the ISM is shown in Fig. 5. Here $\lambda_H$ varies with respect to the energy scale $\mu$ (more precisely with respect to $t = \log\mu$). Different subfigures correspond to different initial values of the other quartic couplings $\lambda_{HS}$ and $\lambda_S$. However, in every case, $\lambda_H$ is fixed to 0.129 at the electroweak scale ($M_Z$), i.e., $\lambda_H = 0.1271$ at one-loop and 0.12604 at two-loops. The one(two)-loop $\beta$-functions are shown by the dotted red (solid blue) curves. Figure 5(a) represents the variation of $\lambda_H$ for the initial values of $\lambda_S = \lambda_{HS} = 0.1$ and for both one- and two-loop, the $\lambda_H$ becomes negative at certain scales, similar to the SM as depicted in Fig. 3. However, as we increase the new coupling values to $\lambda_S = \lambda_{HS} = 0.4$ in Fig. 5(b), we observe an interesting behavior. For one-loop, the evolution of $\lambda_H$ hits a Landau pole at $\mu \approx 10^{10}$ GeV, whereas, for two-loop, we see that $\lambda_H$ attains a Fixed Point around $\mu \approx 10^{15}$ GeV. As we move to Fig. 5(c) for $\lambda_S = \lambda_{HS} = 0.8$, the behavior of LP at one-loop and FP at two-loop appear much earlier, around $\mu \approx 10^6, 10^{10}$ GeV, respectively. Similarly, for $\lambda_S = \lambda_{HS} = 1.2$, the behavior of such LP and FP appear further earlier, around $\mu \approx 10^{5,9}$ GeV, respectively, as shown in Fig. 5(d). The observation of such one-loop LPs and two-loop FPs in ISM is similar to $\phi^4$ theory as described in Sec. II.

Figure 6 shows the behavior of the RG evolution of $\lambda_S$ and $\lambda_{HS}$ where similar observations are found. Figures 6(a) and 6(c) represent the evolution of $\lambda_S$ and $\lambda_{HS}$ with the initial values $\lambda_S(M_Z) = \lambda_{HS}(M_Z) = 0.4$ where the one-loop LP (two-loop FP) are seen at the scales around $10^{10}(10^{16})$ GeV respectively. In Figs. 6(b) and 6(d) for $\lambda_S(M_Z), \lambda_{HS}(M_Z) = 0.8$ we see such LP(FP) at the scales around $10^5(10^9)$ GeV. For larger initial values of $\lambda_S(M_Z), \lambda_{HS}(M_Z)$ the LP(FP) are achieved at much relatively lower energy scales. Even though FP appears $\lesssim 4\pi$ for $\lambda_S$, for $\lambda_{HS}$, it crosses $4\pi$.

Thus, higher initial values of new physics couplings i.e., $\lambda_S, \lambda_{HS}$ promote LPs and FPs for one- and two-loop levels,

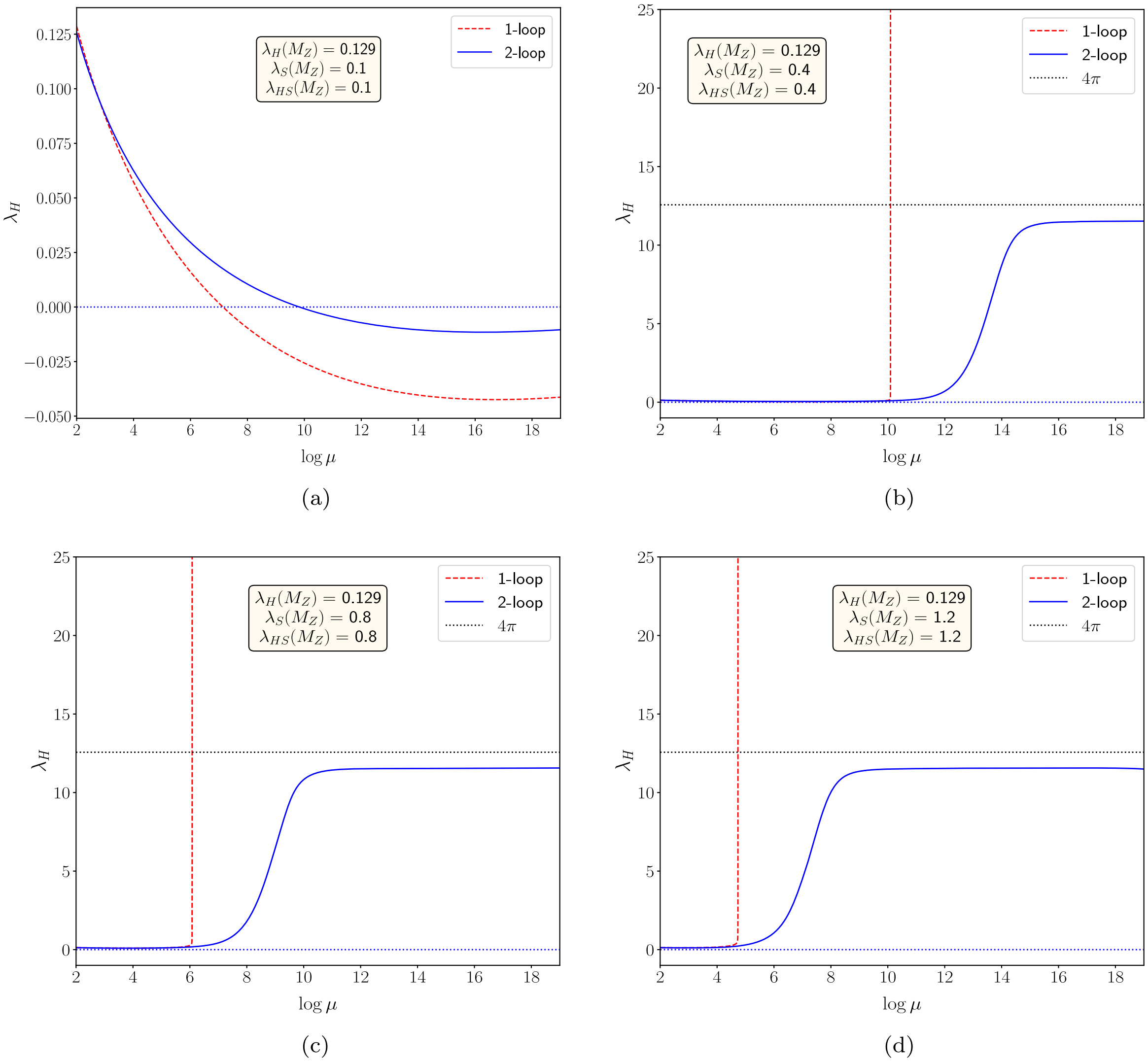

FIG. 5. $\lambda_H - \log \mu$ for ISM at one-loop and two-loop for $\lambda_{HS}(M_Z) = \lambda_S(M_Z)$ for (a)–(d) plots are 0.1, 0.4, 0.8, 1.2, respectively. With $\lambda_H(M_Z) = 0.129$ for all cases.

respectively. For convenience, we define $\alpha = \frac{\lambda_{HS}(M_Z)}{\lambda_S(M_Z)}$, which gives the relative strength of $\Phi^\dagger \Phi S^* S$ term over $(S^* S)^2$ term. Figure 7 gives the values of $\lambda_S$ and $\alpha$ for which there is an FP in the evolution of $\lambda_H(\mu)$ at the two-loop level. The color code gives the energy scale at which the FP begins. The color of the points, as shown in the figure, varies from blue($10^8$ GeV) to yellow ($10^{19}$ GeV). It is evident that for lower initial values of $\lambda_S$, the FPs occur for higher values of $\alpha$. One can see that for $\lambda_S < 1$ the FPs appear around $\mu > 10^{18}$ GeV, and for $\lambda_S > 1.0$ the FPs appear around $\mu > 10^{8\text{–}12}$ GeV. But for very high values of $\lambda_S$ and $\lambda_{HS}$, the FP behavior is absent. In a nutshell, an increase in $\alpha$ promotes FP, provided $\lambda_S$ is small.

If we expand the ISM potential as shown in Eqs. (4.4) and (4.5), we find them as functions of modulus filed terms, without any residual phases, very similar to $\phi^4$ theory in Eq. (2.1). The LP and FP for odd- and even loops are also seen in $\phi^4$ theory [65,66]. These interactions are represented by the tree-level diagrams of the type shown in Fig. 8, where $\phi^a$, $\phi^b = \phi^+$, $\phi^0$, $S$.

$$\lambda_{HS}(\Phi^\dagger \Phi S^* S) + \lambda_S (S^* S)^2 = \lambda_{HS}(|\phi_0|^2 |S|^2 + |\phi^+|^2 |S|^2) + \lambda_S |S|^4, \quad (4.4)$$

$$\lambda_H (\Phi^\dagger \Phi)^2 = \lambda_H (|\phi^0|^4 + 2|\phi^0|^2 |\phi^+|^2 + |\phi^+|^4). \quad (4.5)$$

This seems to be encouraging the appearance of the FPs, as seen in $\phi^4$ theory also. However, the role of the residual phases in spoiling this FP behavior is prominent in the case of IDM, which is discussed in Sec. IV C.

In Fig. 9 we plot the perturbativity limits of $4\pi$. i.e., $\lambda_H(\mu), \lambda_{HS}(\mu), \lambda_S(\mu) < 4\pi$ at one- and two-loop level, along with the other constraints given in Eq. (3.2), in the $\lambda_S(M_Z) - \lambda_{HS}(M_Z)$ plane considering the bounds from perturbative unitarity [60,67,70–74]. Different color codes correspond to different perturbativity scales starting from $10^2$–$10^{19}$ GeV. Generically, the lower the couplings higher the perturbativity scales. At one-loop for the coupling constants $\lambda_S(M_Z) \lesssim 0.2$ and $\lambda_{HS}(M_Z) \lesssim 0.5$, the whole parameter space is perturbative till the Planck scale which is similar even at the two-loop level. However, for larger values of these couplings, the perturbative limits go down to lower scales and the two-loop limits also differ from one-loop. The one-loop perturbativity limits as shown in Fig. 9(a), when $\lambda_S(M_Z) > 1.5$, regardless of the value of $\lambda_{HS}$, perturbativity hits at energy scales between $10^2$–$10^3$ GeV. For relatively higher values of $\lambda_{HS}(M_Z) > 2.0$, relatively lower values of $\lambda_S$ hit the perturbativity limit for the scale $10^2$–$10^3$ GeV. When $1.0 < \lambda_{HS} < 1.5$, the perturbativity hits around $10^3$–$10^4$ GeV. For the two-loop, the perturbative limit as shown in Fig. 9(b), gets relaxed and larger values of quartic couplings are now allowed. For $\lambda_S(M_Z) > 3.0$, it hits the perturbativity for the scale $10^2$–$10^3$ GeV. For larger values of $\lambda_{HS}(M_Z) > 2.8$, the perturbativity hits at a similar scale for lower $\lambda_S(M_Z)$. The yellow color region inside the black contour shows the unstable points that fail to satisfy the bounded-from-below stability conditions in Eq. (4.2). For multi-TeV dark matter, where these parameter regions are allowed by dark matter, these theoretical constraints are more relevant [33,41,79].

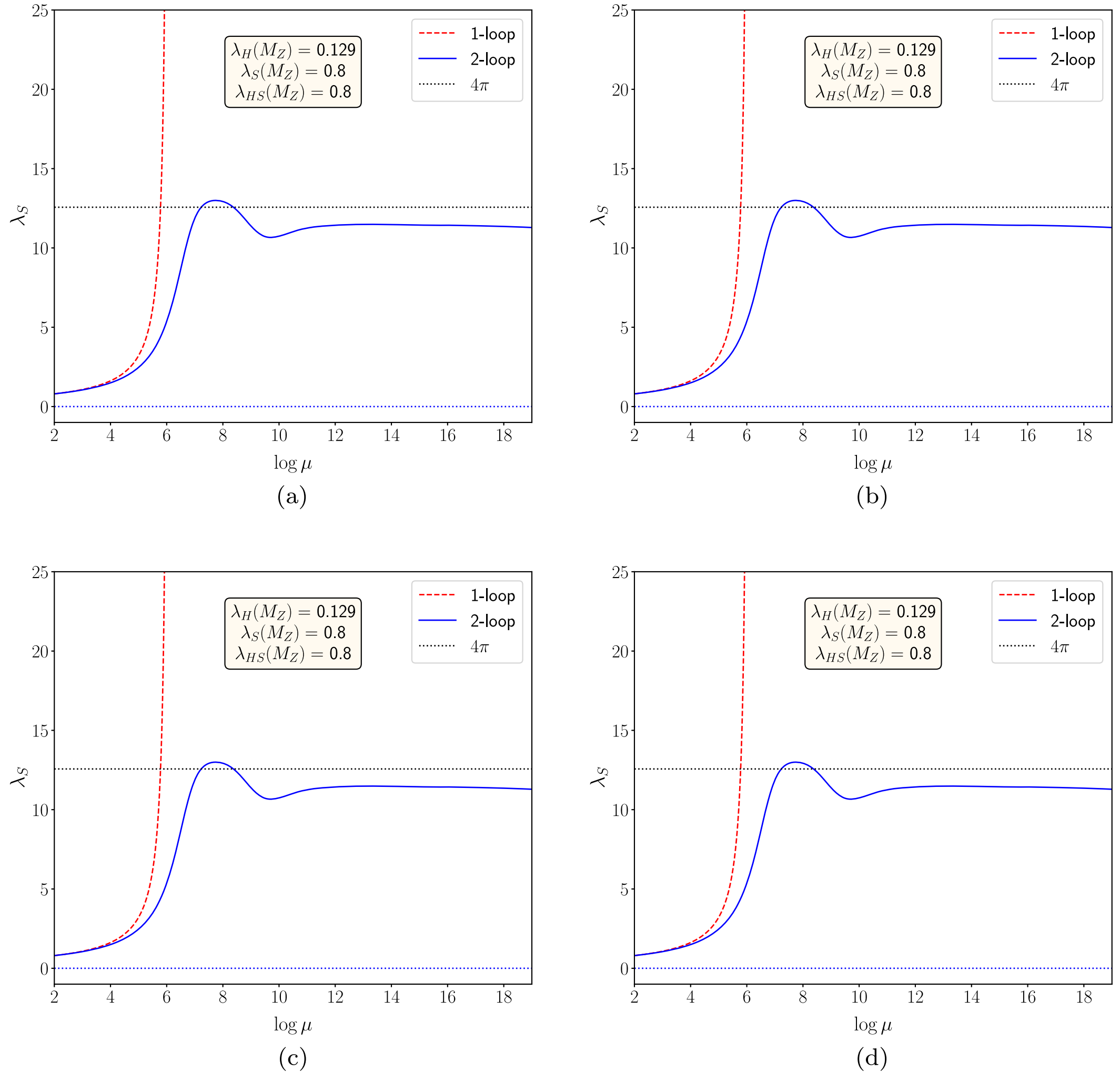

FIG. 6. $\lambda_{HS} - \log \mu$ in (a), (b) and $\lambda_S - \log \mu$ in (c), (d) are depicted for ISM at one-loop and two-loop for $\lambda_H(M_Z) = 0.129$. $\lambda_{HS}(M_Z) = \lambda_S(M_Z) = 0.4$, 0.8 are fixed for (a), (c), and (b), (d), respectively.

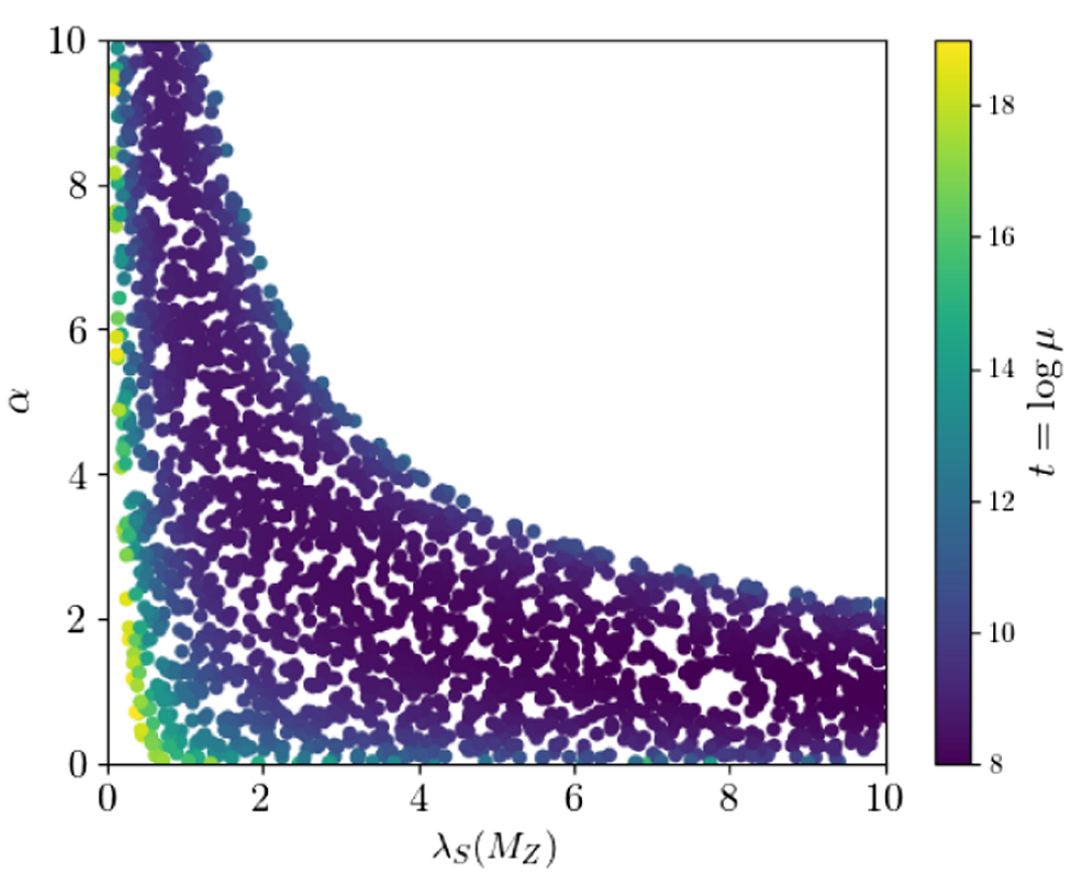

FIG. 7. Occurrence of FP at two-loop level for different initial values of $\alpha = \frac{\lambda_{HS}(M_Z)}{\lambda_S(M_Z)}$ and $\lambda_S(M_Z)$. Initial value of $\lambda_H(M_Z) = 0.129$ for all the points.

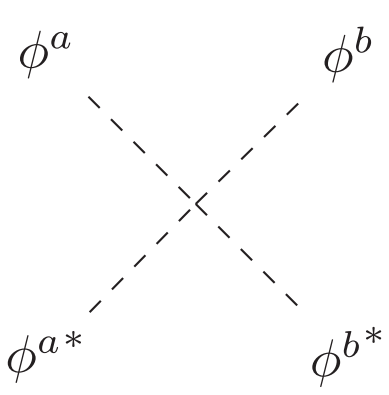

FIG. 8. Type of scalar four-point vertices in Inert Singlet Model.

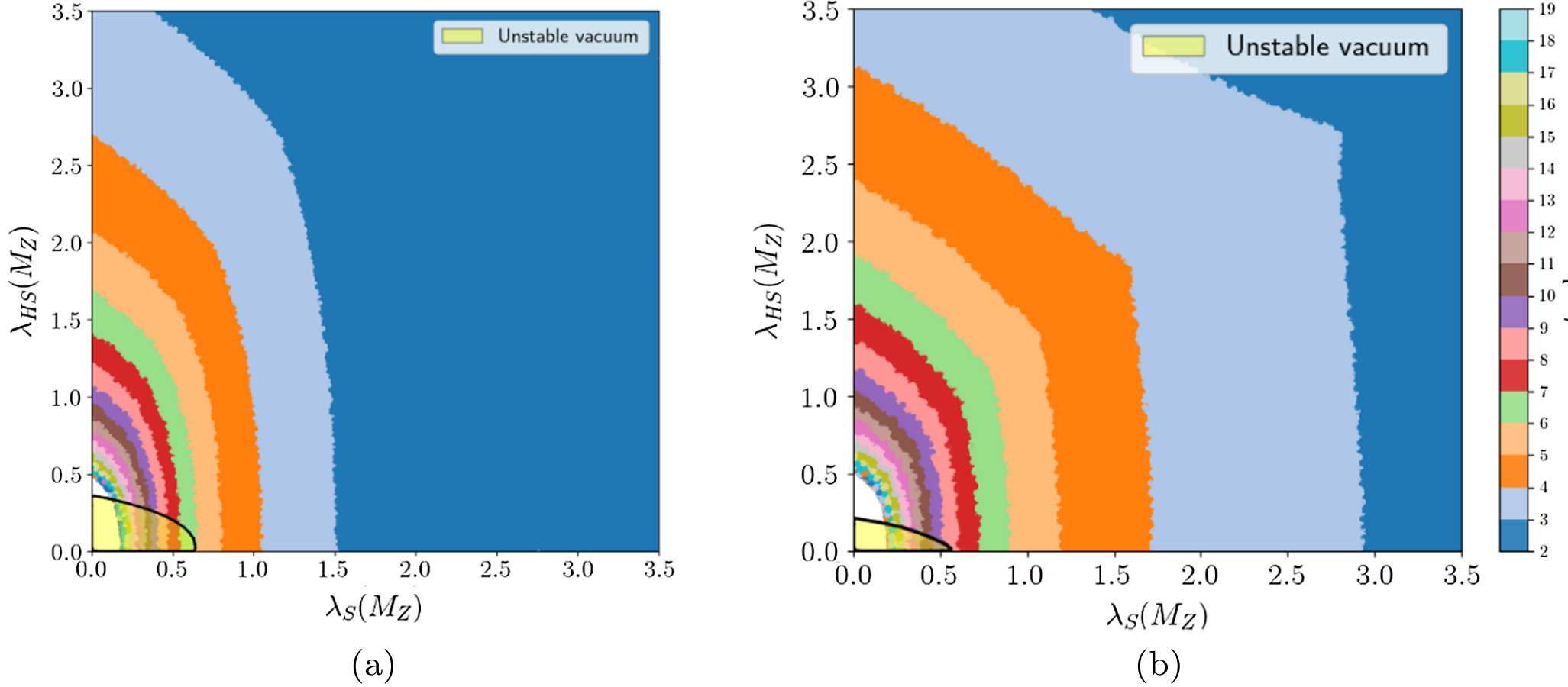

FIG. 9. Perturbativity limit of RGE evolution of Inert Singlet Model at one-loop level in (a) and two-loop level in (b). Initial value of Higgs-self quartic coupling $\lambda_H(M_Z) = 0.129$ for all cases. $\lambda_S(M_Z)$ is along the x-axis and $\lambda_{HS}(M_Z)$ is along the y-axis.

### B. Inert Triplet Model

The Inert Triplet Model (ITM) consists of an $SU(2)_L$ triplet scalar $T$ with zero hypercharge ($Y = 0$) along with the SM Higgs doublet. The triplet as given in Eq. (4.6) is odd under a $\mathbb{Z}_2$ symmetry and thus does not play any role in EWSB, therefore it is called inert. The neutral component provides a real scalar dark matter [16,25,26,37–41].

$$\Phi = \begin{pmatrix} \phi^+ \\ \phi^0 \end{pmatrix} \quad \text{and} \quad T = \frac{1}{2}\begin{pmatrix} T^0 & \sqrt{2}T^+ \\ \sqrt{2}T^- & -T^0 \end{pmatrix}. \tag{4.6}$$

The triplet field is odd under $\mathbb{Z}_2$ symmetry while all the SM fields are even under $\mathbb{Z}_2$. The covariant derivative for the triplet is given by,

$$D_\mu T = \partial_\mu T - i g_2 \left[\frac{\sigma^a}{2} W_\mu^a T\right], \tag{4.7}$$

while the scalar potential is as follows:

$$V = m_H^2 \Phi^\dagger \Phi + m_T^2 \mathrm{Tr}[T^\dagger T] + \lambda_H (\Phi^\dagger \Phi)^2 + \lambda_T (\mathrm{Tr}[T^\dagger T])^2 + \lambda_{HT} (\Phi^\dagger \Phi) \mathrm{Tr}[T^\dagger T], \tag{4.8}$$

where $\lambda_H$ is the SM-like Higgs self-coupling, $\lambda_T$ is the triplet self-coupling and $\lambda_{HT}$ is the Higgs-triplet portal coupling. Here the triplet and doublet do not mix and $T^\pm, T^0$ are also the triplet mass eigenstates. For a stable vacuum, the quartic coupling must satisfy the following stability conditions [32,41,42]

$$\lambda_H \geq 0, \qquad \lambda_T \geq 0, \qquad \lambda_{HT}^2 - 4\lambda_H \lambda_T \geq 0. \tag{4.9}$$

The $\beta$-function of SM-like Higgs quartic coupling generated using SARAH 4.15.1 [78] is given, at the one-loop level, in Eq. (4.10) and the rest of the $\beta$-functions in two-loop are given in Appendix A 3. Even though the potential and new physics couplings look similar to ISM, the $\beta$-function looks very different from the $\beta_{\lambda_H}^{\mathrm{ISM}}$ in the case of ISM [Eq. (4.3)] since the triplet is in a nontrivial representation of $SU(2)$ that involves the gauge coupling $g_2$.

$$\beta_{\lambda_H}^{\mathrm{ITM}} = \beta_{\lambda_H}^{\mathrm{SM}} + \frac{1}{(4\pi)^2}\left(\frac{3}{2}\lambda_{HT}^2\right) + \frac{1}{(4\pi)^4} \times \left(\frac{11}{2} g_2^4 \lambda_H + \frac{5}{4} g_2^4 \lambda_{HT} + \frac{9}{2} g_2^2 \lambda_{HT}^2 - 15 \lambda_H \lambda_{HT}^2 - 2\lambda_{HT}^3 - \frac{7}{20} g_1^2 g_2^4 - \frac{7}{4} g_2^6\right). \tag{4.10}$$

The evolution of SM-like Higgs quartic coupling $\lambda_H$ is shown in Fig. 10. Different figures correspond to different initial values of the other quartic couplings, say $\lambda_T$ and $\lambda_{HT}$ where the initial value of $\lambda_H(M_Z) = 0.129$ at the electro-weak scale. Figure 10(a) shows the variation of $\lambda_H$ for lower initial values of $\lambda_{HT}(M_Z) = \lambda_T(M_Z) = 0.1$. The results look similar to SM and ISM as both one(two)-loop $\lambda_H$ go negative at certain scales. As we increase $\lambda_{HT}(M_Z) = \lambda_T(M_Z) = 0.4$ in Fig. 10(b), $\lambda_H$ becomes nonzero till the Planck scale for both one(two)-loop $\beta_{\lambda_H}$. However, unlike Fig. 5(b), here we do not get any Landau poles at one-loop or Fixed Points at two-loop for $\lambda_{HT}(M_Z) = \lambda_T(M_Z) = 0.4$.

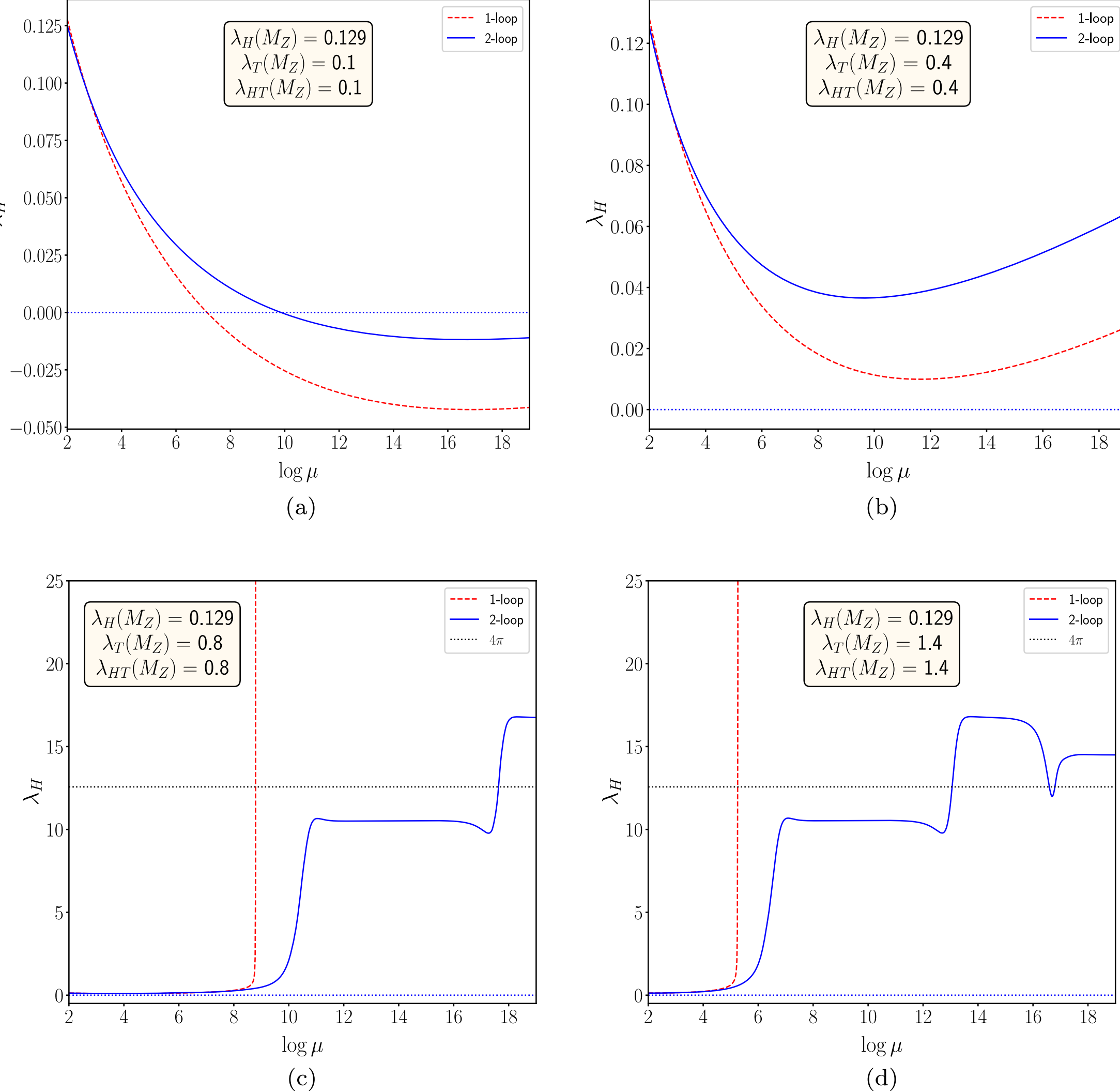

FIG. 10. One-loop and two-loop RGE evolution of SM Higgs-like quartic coupling in the Inert Triplet Model. Initial value of $\lambda_H(M_Z) = 0.129$ for all cases. Initial values of $\lambda_{HT}(M_Z)$ and $\lambda_T(M_Z)$ for (a), (b), (c), (d) plots are 0.1, 0.4, 0.8, 1.4, respectively.

From Fig. 10(c) it is clear that there is an FP behavior in the two-loop evolution of $\lambda_H$ (solid blue curve) for the initial value of $\lambda_{HT}(M_Z) = \lambda_T(M_Z) = 0.8$. We get a Landau pole in the one-loop evolution (dotted red curve). Similar behavior can be found in Fig. 10(d) for the initial value of $\lambda_{HT}(M_Z) = \lambda_T(M_Z) = 1.2$ also. Similar to the case of ISM, the higher initial values will lead to attaining the LP and FP at relatively lower energy scales. However, we notice the appearance of two different Fixed Points for two different values of $\lambda_H$, one below $4\pi$ and one above $4\pi$, which was absent in $\phi^4$ theory and ISM.

To complete the analysis of scalar quartic couplings we also see the evolution of $\lambda_T$ and $\lambda_{HT}$ verses the scale in Figs. 11(a) and 11(b) respectively for the initial values of $\lambda_T(M_Z) = \lambda_{HT}(M_Z) = 0.8$. Though we notice FP behavior in both the cases, however they appear above $4\pi$ in the coupling constant values. Though the appearance of the FP happens in the scale $\sim 10^{10}$ GeV, similar to $\lambda_H$, around $\mu \gtrsim 10^{18}$ GeV, we see a dip in the couplings that go beyond zero.

One could obtain the parameter ranges where FP is occurring for different initial values of $\lambda_T(M_Z)$ and $\lambda_{HT}(M_Z)$ for a fixed value of $\lambda_H(M_Z)$. For convenience, we define $\alpha = \frac{\lambda_{HT}(M_Z)}{\lambda_S(M_Z)}$, which gives us the relative strength of $\Phi^\dagger \Phi S^* S$ term over $[\mathrm{Tr}(T^\dagger T)]^2$ term.

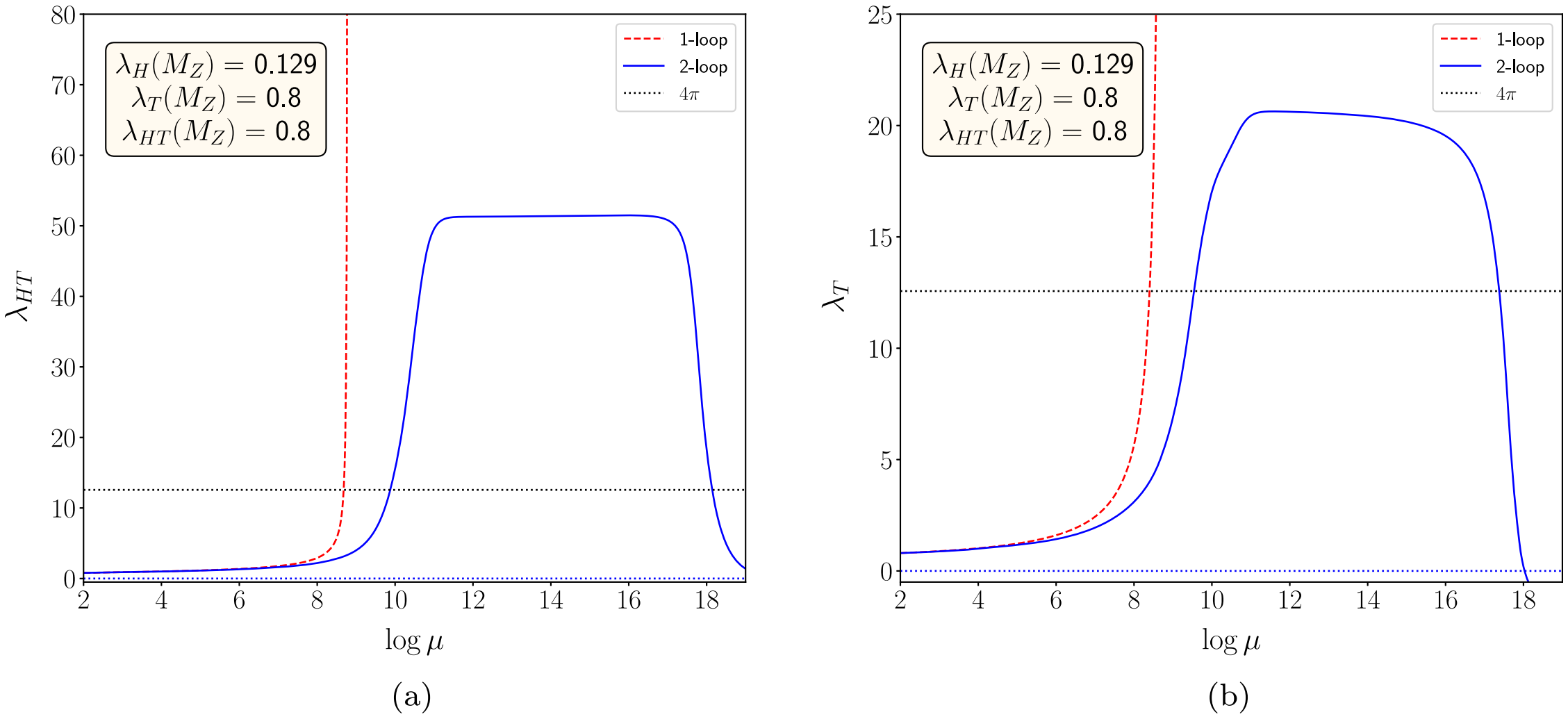

FIG. 11. $\lambda_{HT} - \log \mu$ in (a), and $\lambda_T - \log \mu$ in (b) are depicted for ITM at one-loop and two-loop for $\lambda_H(M_Z) = 0.129$, $\lambda_{HT}(M_Z) = \lambda_T(M_Z) = 0.8$.

Figure 12 gives the values of $\lambda_T(M_Z)$ and $\alpha$ for which there are two-loop FPs in the evolution of $\lambda_H(\mu)$. The color code gives the energy scale at which the FP begins. For lower initial values of $\lambda_T(M_Z)$, the FP occurs when the $\alpha$ value is higher. As initial values of $\lambda_T(M_Z)$ and $\alpha$ become higher, the FP occurs at a lower energy scale. But for very high values of $\lambda_T(M_Z)$ and $\lambda_{HT}(M_Z)$, the FP behavior is absent. In conclusion, the increase in the $\alpha$ promotes the FP, provided $\lambda_T(M_Z)$ is small as in the case of ISM.

As we discussed in the case of the ISM, we like to see if the scalar potential depends only on the modulus-squared terms and not on the phases of the complex fields. The $\lambda_H$ term is given by,

$$\lambda_H(\Phi^\dagger\Phi)^2 = \lambda_H(|\phi^0|^4 + 2|\phi^0|^2|\phi^+|^2 + |\phi^+|^4). \quad (4.11)$$

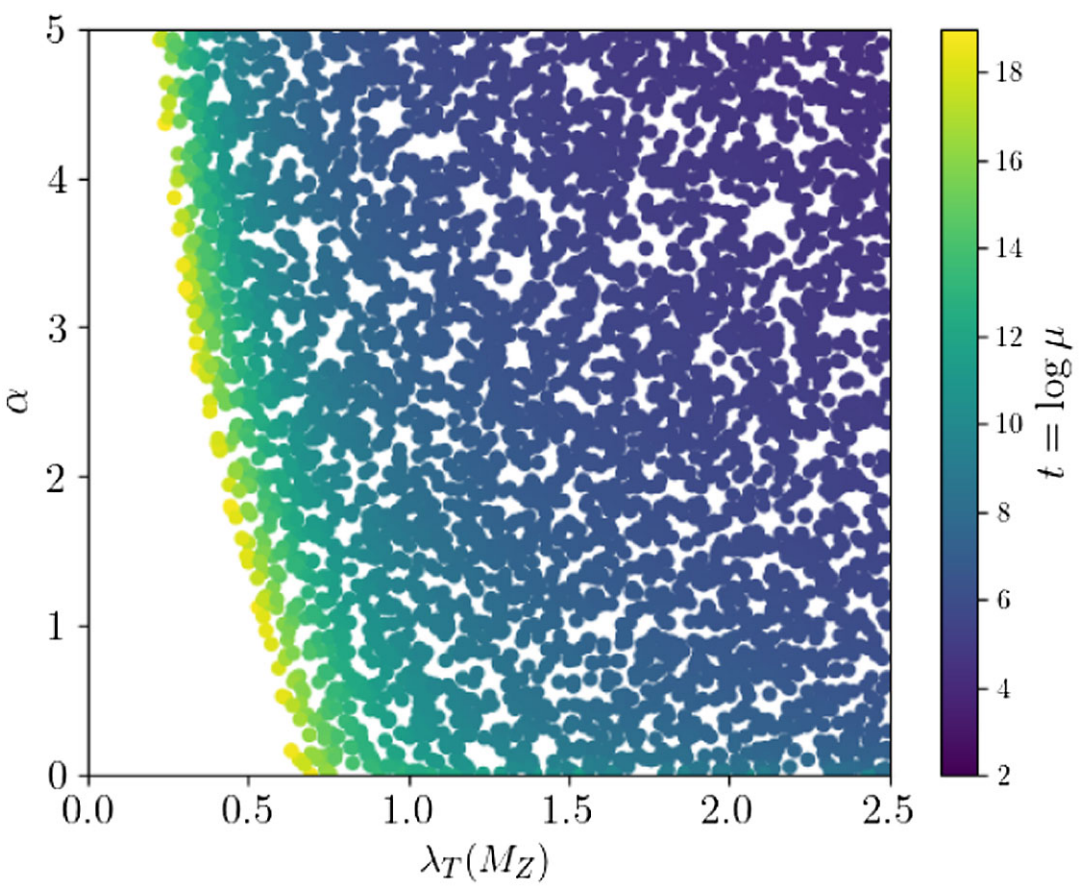

FIG. 12. Occurrence of two-loop FP for different initial values of $\alpha = \frac{\lambda_{HT}(M_Z)}{\lambda_T(M_Z)}$ and $\lambda_T(M_Z)$. Initial value of $\lambda_H(M_Z) = 0.129$ for all points.

The $\lambda_T$ term is given by,

$$\lambda_T(\mathrm{Tr}(T^\dagger T))^2 = \lambda_T\left(\frac{1}{4}|T^0|^4 + \frac{1}{2}|T^0|^2|T^-|^2 + \frac{1}{4}|T^-|^4 + \frac{1}{2}|T^0|^2|T^+|^2 + \frac{1}{2}|T^-|^2|T^+|^2 + \frac{1}{4}|T^+|^4\right). \quad (4.12)$$

The $\lambda_{HT}$ term is given by,

$$\lambda_{HT}(\Phi^\dagger\Phi)\mathrm{Tr}(T^\dagger T) = \frac{1}{2}\lambda_{HT}(|T^0|^2|\phi^0|^2 + |T^-|^2|\phi^0|^2 + |T^+|^2|\phi^0|^2 + |T^0|^2|\phi^+|^2 + |T^-|^2|\phi^+|^2 + |T^+|^2|\phi^+|^2). \quad (4.13)$$

Like in the case of ISM, all the terms in the scalar potential depend only on the absolute values of the fields and not on the residual phases. Later in Sec. IV C, we will illustrate that these are the kind of terms that favor a two-loop FP behavior.

Figure 13 presents the perturbative limit of $4\pi$, i.e., $\lambda_H(\mu), \lambda_{HT}(\mu), \lambda_T(\mu) < 4\pi$ at one- and two-loop level, along with the other constraints given in Eq. (3.2), in the $\lambda_T(M_Z) - \lambda_{HT}(M_Z)$ plane for different scales similar to Fig. 9. Planck scale perturbativity at one-loop can be achieved till $\lambda_T(M_Z) \lesssim 0.5, \lambda_{HT}(M_Z) \lesssim 0.4$, and further increment of $\lambda_{HT}(M_Z)$ lowers the perturbative scale. For two-loop, a similar observation is made for $\lambda_T(M_Z) \lesssim 0.6, \lambda_{HT}(M_Z) \lesssim 0.4$. For higher values of couplings, the perturbativity scales decrease further and the two-loop limits differ more from the one-loop. In Fig. 13(a), at the one-loop level, we see that for $\lambda_T(M_Z) > 2.0$ the perturbativity scale appears at $10^{2-3}$ GeV. As we increase the $\lambda_{HT}(M_Z)$ the perturbative limit happens for even lower values of $\lambda_T(M_Z)$ at the same scale. For two-loop as described in Fig. 13(b), when $\lambda_T(M_Z) > 2.6$ the perturbativity scale appears at $10^{2\text{–}3}$ GeV As we increase the $\lambda_{HT}(M_Z)$ further the allowed values of $\lambda_T(M_Z)$ reduce further for the same scale. Overall the two-loop results are more relaxed as compared to one-loop, similar to the inert singlet case. Black-contoured yellow regions correspond to an unstable vacuum which fails to satisfy Eq. (4.9). Generically, this happens for lower values of the quartic couplings.

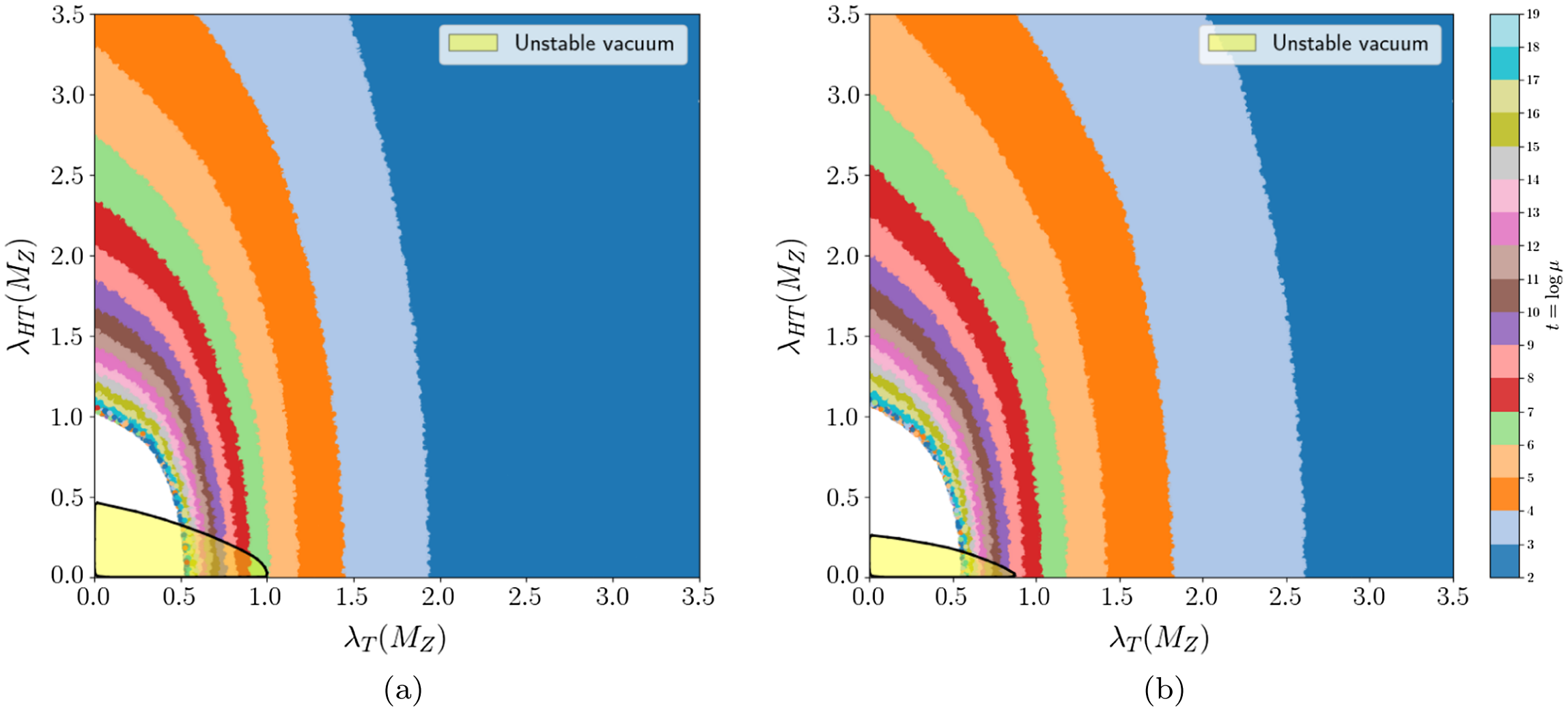

FIG. 13. Perturbativity limit of RGE evolution of Inert Triplet Model at one-loop level in (a) and two-loop level in (b). Initial value of Higgs-self quartic coupling $\lambda_H(M_Z) = 0.129$ for all cases. $\lambda_T(M_Z)$ is along the x-axis and $\lambda_{HT}(M_Z)$ is along the y-axis.

### C. Inert Doublet Model

In the Inert Doublet Model (IDM), the SM Higgs sector is augmented with another $SU(2)_L$ doublet ($\Phi_2$) that has the same hypercharge as the SM Higgs field ($\Phi_1$), $Y = \frac{1}{2}$. This doublet is odd under $\mathbb{Z}_2$ symmetry while all the SM fields are even [18–32,34,37]. The scalar potential of the IDM is written as,

$$
\begin{aligned}
V(\Phi_1, \Phi_2) = {} & m_{11}^2 \Phi_1^\dagger \Phi_1 + m_{22}^2 \Phi_2^\dagger \Phi_2 + \lambda_1 (\Phi_1^\dagger \Phi_1)^2 \\
& + \lambda_2 (\Phi_2^\dagger \Phi_2)^2 + \lambda_3 (\Phi_1^\dagger \Phi_1)(\Phi_2^\dagger \Phi_2) \\
& + \lambda_4 (\Phi_1^\dagger \Phi_2)(\Phi_2^\dagger \Phi_1) + \lambda_5 ((\Phi_1^\dagger \Phi_2)^2 + \text{H.c.}),
\end{aligned}
\tag{4.14}
$$

where $\Phi_1 = (\,\phi_1^+ \phi_1^0\,)^T$ and $\Phi_2 = (\,\phi_2^+ \phi_2^0\,)^T$. The field $\Phi_2$ is called inert because it does not take part in EWSB. The coefficients $m_{11}^2$ and $m_{22}^2$ are the mass terms, $\lambda_1$ is the SM-like Higgs self-coupling (which is equivalent to $\lambda_H$ in ISM and ITM), $\lambda_2$ is the inert doublet self-coupling, and $\lambda_{3,4,5}$ are the different portal couplings. This gives us two $CP$-even $h$, $H$, one $CP$-odd $A^0$, and one charged Higgs boson $H^\pm$ as physical mass eigenstates [32,33]. The stability conditions for this potential are the following [18,32,49]

$$
\begin{aligned}
& \lambda_1 \geq 0, \qquad \lambda_2 \geq 0, \qquad \lambda_3 \geq -2\sqrt{\lambda_1 \lambda_2}, \\
& \lambda_3 + \lambda_4 - 2|\lambda_5| \geq -2\sqrt{\lambda_1 \lambda_2}.
\end{aligned}
\tag{4.15}
$$

The $\beta$-functions for this model are also obtained using SARAH 4.15.1. The $\beta_{\lambda_1}$ at one-loop is shown in Eq. (4.16) and the complete set of two-loop $\beta$-functions are given in Appendix A 4.

$$
\beta_{\lambda_1}^{\text{IDM}} = \beta_{\lambda_1}^{\text{SM}} + \frac{1}{(4\pi)^2} (2\lambda_3^2 + 2\lambda_3 \lambda_4 + \lambda_4^2 + 4\lambda_5^2)
\tag{4.16}
$$

The evolutions of SM-like Higgs quartic coupling $\lambda_1$ in the inert doublet scenario, for different initial values of the other quartic couplings $\lambda_{i\neq1}$, are shown in Fig. 14. Like in the cases of ISM and ITM, the one-loop evolution is shown in the red dashed curve and the two-loop evolution is the blue solid curve. Similar to the previous case, we take $\lambda_H(M_Z) = 0.129$ and other quartic couplings are varied as $\lambda_{i\neq1}(M_Z) = 0.01$, 0.4, 0.8, and 1.2 for Figs. 14(a)–14(d), respectively. We can see from Fig. 14(a) that $\lambda_1 = \lambda_H$ becomes negative for both one (two)-loop as in SM. However, it is worth mentioning that when we increase $\lambda_{i\neq1}(M_Z) = 0.1$, the one-loop will remain negative at higher scales but in the two-loop, it remains positive till the Planck scale. When we increase the new physics couplings $\lambda_{i\neq1}(M_Z) = 0.4$, $\lambda_1$ hits LP at $\mu \approx 10^{7.5} (10^{8.7})$ GeV for one(two)-loop $\beta$-functions, respectively. The behavior is similar for higher values of $\lambda_{i\neq1}$ i.e., $\lambda_{i\neq1}(M_Z) = 0.8$ where the LPs appear much earlier $\mu \approx 10^{4.7}(10^{5.7})$ GeV for one(two)-loop $\beta$-functions, respectively as can be seen from Fig. 14(c). Finally, in Fig. 14(d) we plot for $\lambda_{i\neq 1}(M_Z) = 1.2$ where the LPs come around $\mu \approx 10^{3.6}(10^{4.5})$ GeV for one(two)-loop $\beta$-functions, respectively. Even though we do not find any FPs for higher values of $\lambda_{i\neq 1}$, unlike SM or ISM, the situation is far more interesting as we elaborate in the following paragraphs.

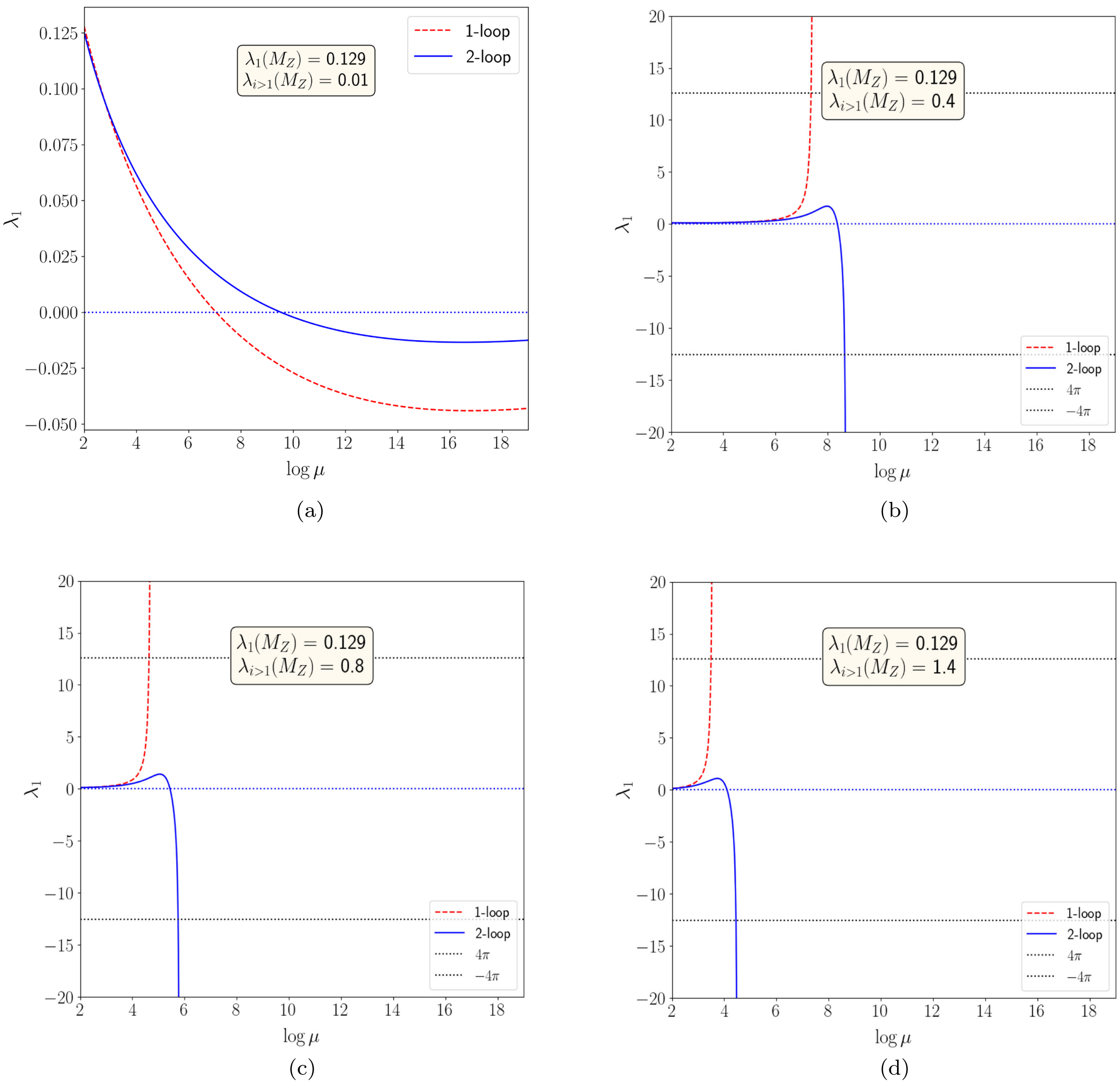

FIG. 14. Two-loop RGE evolution of SM Higgs-like quartic coupling in Inert Doublet Model. The initial value of $\lambda_{i\neq 1}(M_Z)$ for (a), (b), (c), (d) plots are 0.01, 0.4, 0.8, 1.4, respectively. Initial value of Higgs-self quartic coupling $\lambda_1(M_Z) = 0.129$ for all cases.

Nonappearance of the FP in the case of IDM can be explained if we analyze the portal couplings in detail, as we did for ISM and ITM in Secs. IV A and IV B, respectively. It can be seen from Eq. (4.17), that $\lambda_{1,2,3}$-terms are terms that do not contain any residual phases. They are similar to the terms obtained for ISM in Eqs. (4.4), (4.5), and for ITM in Eqs. (4.12) and (4.13) in the sense that they only depend on the square of the absolute values of the fields.

$$
\begin{aligned}
\lambda_1(\Phi_1^\dagger\Phi_1)^2 &= \lambda_1(|\phi_1^0|^4 + 2|\phi_1^0|^2|\phi_1^+|^2 + |\phi_1^+|^4),\\
\lambda_2(\Phi_2^\dagger\Phi_2)^2 &= \lambda_2(|\phi_2^0|^4 + 2|\phi_2^0|^2|\phi_2^+|^2 + |\phi_2^+|^4),\\
\lambda_3(\Phi_1^\dagger\Phi_1)(\Phi_2^\dagger\Phi_2) &= \lambda_3(|\phi_1^0|^2|\phi_2^0|^2 + |\phi_1^+|^2|\phi_2^0|^2 + |\phi_1^0|^2|\phi_2^+|^2 + |\phi_1^+|^2|\phi_2^+|^2).
\end{aligned}
\tag{4.17}
$$

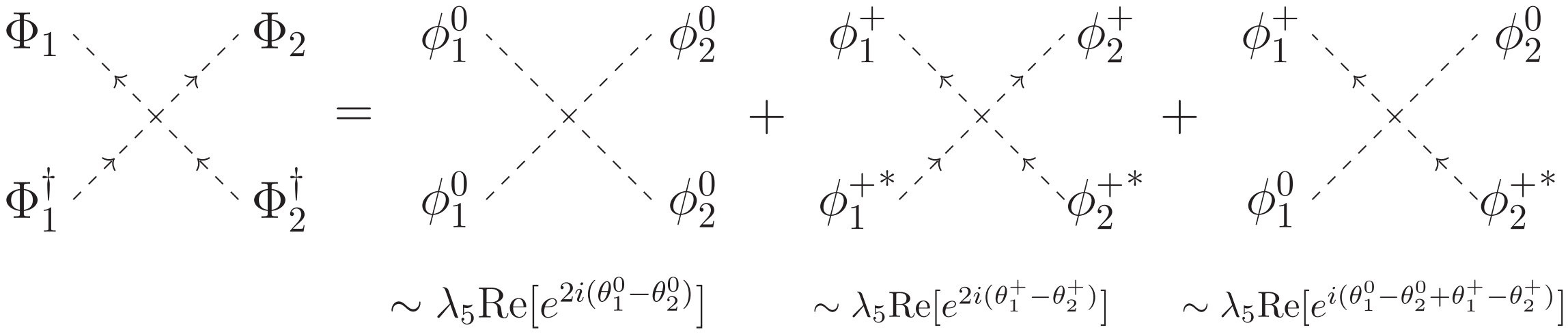

FIG. 15. Relative residual phases associated to $\lambda_5$ term.

(a)

(b)

(c)

(d)

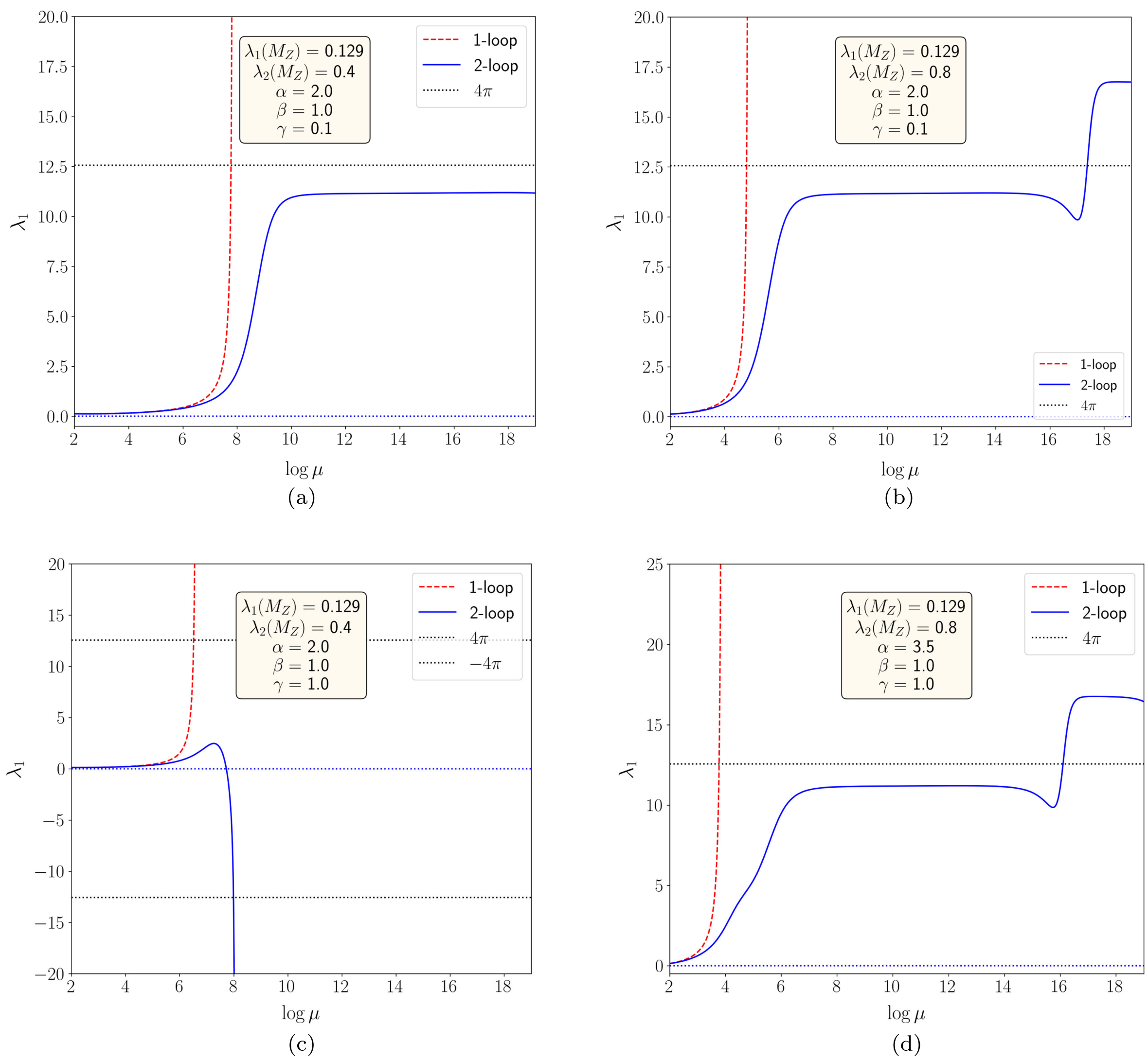

FIG. 16. Two-loop RGE evolution of SM Higgs-like quartic coupling in Inert Doublet Model for different fixed values of $\lambda_{i\neq 1}$. Initial value of $\lambda_H = 0.129$ for all cases. The initial values of $(\lambda_2, \alpha, \beta, \gamma)$ for (a), (b), (c), and (d) are (0.4, 2.0, 1.0, 0.1), (0.8, 2.0, 1.0, 0.1), (0.4, 2.0, 1.0, 1.0), and (0.8, 3.5, 1.0, 1.0), respectively.

However, if we expand the $\lambda_{4,5}$ terms, as given in Eqs. (4.18) and (4.19), we can see that they not only depend on the modulus of the terms but also depend on the residual phases.

$$\lambda_4(\Phi_1^\dagger\Phi_2)(\Phi_2^\dagger\Phi_1) = \lambda_4(|\phi_1^0|^2|\phi_2^0|^2 + \phi_1^0\phi_2^+\phi_1^{+*}\phi_2^{0*} + \phi_1^+\phi_2^0\phi_1^{0*}\phi_2^{+*} + |\phi_1^+|^2|\phi_2^+|^2), \tag{4.18}$$

$$\lambda_5((\Phi_1^\dagger\Phi_2)^2 + \mathrm{H.c.}) = \lambda_5(\phi_2^{02}\phi_1^{0*2} + 2\phi_2^0\phi_2^+\phi_1^{0*}\phi_1^{+*} + \phi_2^{+2}\phi_1^{+*2} + \phi_1^{02}\phi_2^{0*2} + 2\phi_1^0\phi_1^+\phi_2^{0*}\phi_2^{+*} + \phi_1^{+2}\phi_2^{+*2}). \tag{4.19}$$

This point becomes more obvious if we expand those terms with the following form of the fields.

$$\phi_1^+ = |\phi_1^+|e^{i\theta_1^+} \quad \phi_1^0 = |\phi_1^0|e^{i\theta_1^0} \quad \phi_2^+ = |\phi_2^+|e^{i\theta_2^+} \quad \phi_2^0 = |\phi_2^0|e^{i\theta_2^0}, \tag{4.20}$$

where $\{\theta_1^+, \theta_1^0, \theta_2^+, \theta_2^0\} \in \mathbb{R}$, are the real-valued phases of the corresponding complex fields. In Fig. 15 we show the phases associated with the $\lambda_5$ term.

Upon expansion the $\lambda_{4,5}$ terms become,

$$\lambda_4(\Phi_1^\dagger\Phi_2\Phi_2^\dagger\Phi_1) = \lambda_4|\phi_1^0\phi_2^0|^2 + \lambda_4|\phi_1^+\phi_2^+|^2 + 2\lambda_4|\phi_1^0\phi_1^+\phi_2^0\phi_2^+|\cos(\theta_1^0 - \theta_1^+ - \theta_2^0 + \theta_2^+) \tag{4.21}$$

$$\begin{aligned}\lambda_5((\Phi_1^\dagger\Phi_2)^2 + \mathrm{H.c.}) = {}& \lambda_5 2|\phi_1^0|^2|\phi_2^0|^2\cos(2\theta_1^0 - 2\theta_2^0) + \lambda_5 2|\phi_1^+|^2|\phi_2^+|^2\cos(2\theta_1^+ - 2\theta_2^+) \\ & + 4\lambda_5|\phi_1^0\phi_1^+\phi_2^0\phi_2^+|\cos(\theta_1^0 + \theta_1^+ - \theta_2^0 - \theta_2^+).\end{aligned} \tag{4.22}$$

Even though these terms are real-valued, we can see that they depend not only on the squares of the absolute values of the complex fields but also on the residual phases associated with these fields. It is also clear that only one out of the three terms in the expansion of $\lambda_4(\Phi_1^\dagger\Phi_2)(\Phi_2^\dagger\Phi_1)$ has these residual phases but all the three terms in the expansion of $\lambda_5((\Phi_1^\dagger\Phi_2)^2 + \mathrm{H.c.})$ have such phases.

We see that the terms associated with the residual phases, i.e., $\lambda_{4,5}$ terms, drift away from the $\beta$-function of $\lambda_1 = \lambda_H$ to attain a Fixed Point. On the other hand, the $\lambda_{1,2,3}$ terms (the term without any such residual phases, i.e., the modulus-squared terms) enhance the probability of getting an FP in the $\lambda_H$ evolution. This behavior was seen in the case of ISM in Sec. IV A, where the presence of additional scalar terms like $\lambda_{S,HS}$ and the absence of terms with residual phase enhances such possibility of FPs in the composition of ISM. A similar observation is also noted for ITM in Sec. IV B.

The appearance of FPs in a scalar field theory may be attributed to the absence of any residual phases from the scalar potential which is true for the $O(N)$ symmetric potential in a $\phi^4$ theory [65,66], which is also realized in the case of ISM and ITM. However, this property is lost in IDM due to the existence of $\lambda_{4,5}$ couplings. To show this is the case, we define $\alpha = \frac{\lambda_3(M_Z)}{\lambda_2(M_Z)}$, which enhances the possibility of the two-loop FP, and the damage to the FPs comes from $\lambda_{4,5}$ terms; we define their relative strengths, with respect to $\lambda_2$-term, as $\beta = \frac{\lambda_4(M_Z)}{\lambda_2(M_Z)}$ and $\gamma = \frac{\lambda_5(M_Z)}{\lambda_2(M_Z)}$.

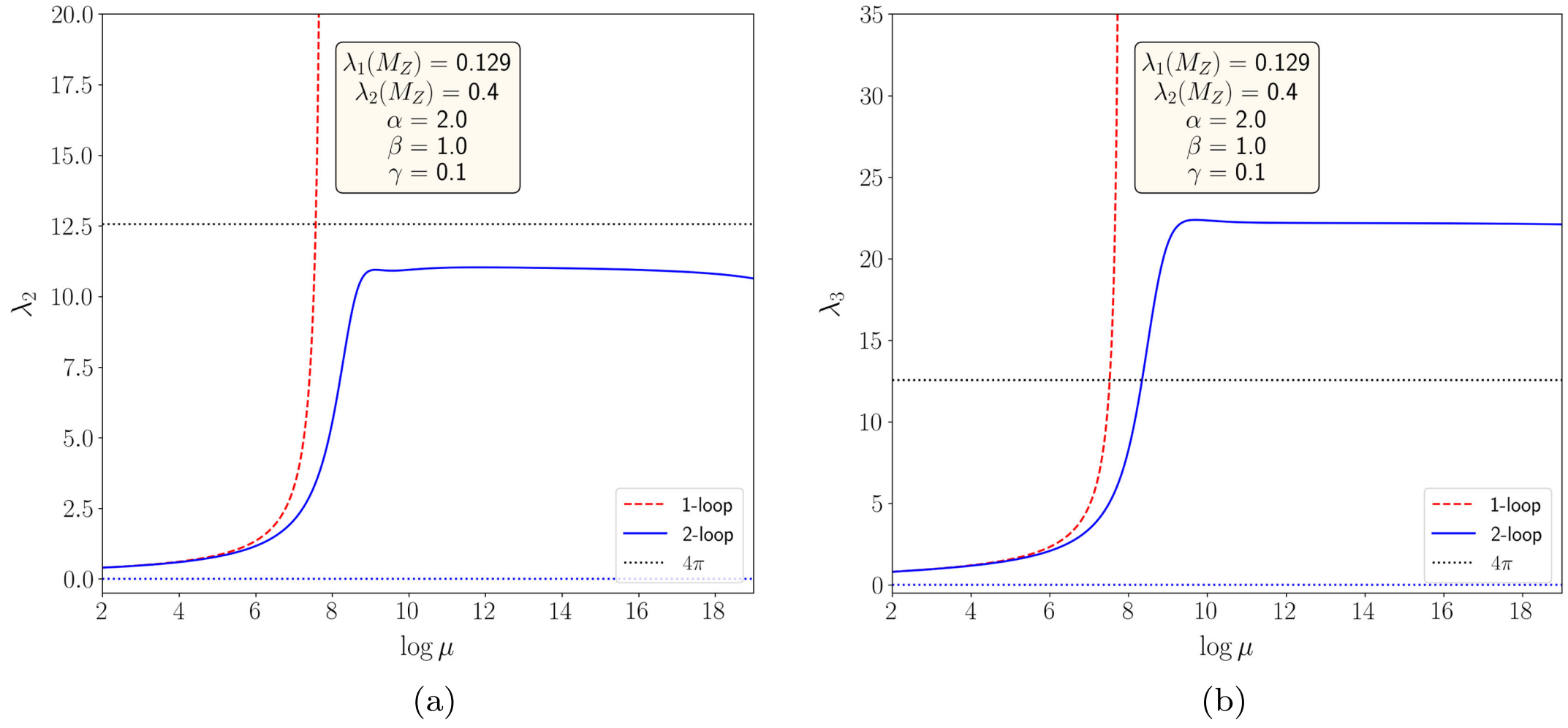

FIG. 17. Two-loop RGE evolution of $\lambda_2$ and $\lambda_3$ in Inert Doublet Model for different fixed values of $\lambda_{i\neq1}$. Initial value of $\lambda_H = 0.129$ for all cases. The initial values of $(\lambda_2, \alpha, \beta, \gamma)$ for (a), and (b) is (0.4, 2.0, 1.0, 0.1), respectively.

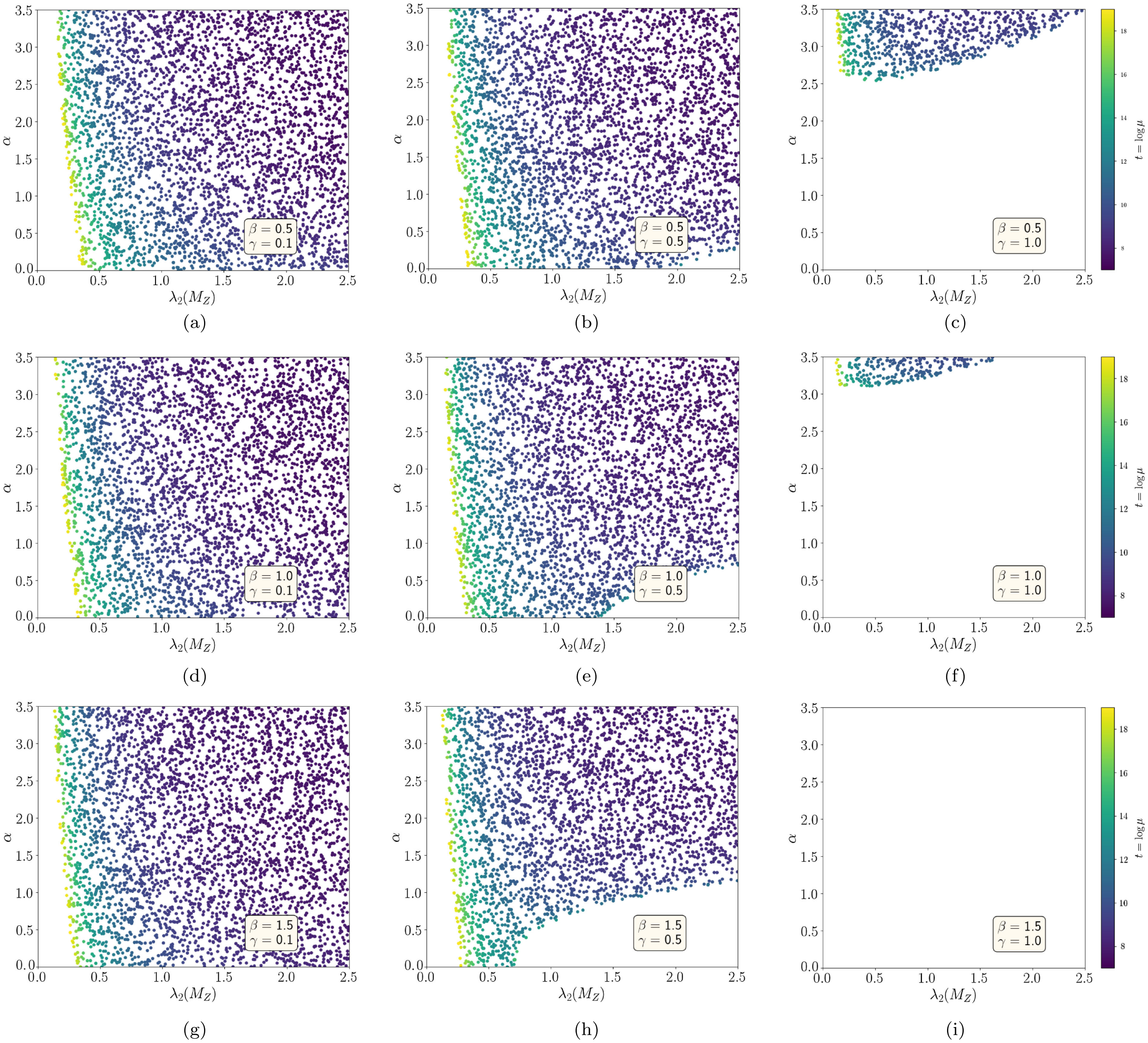

FIG. 18. Occurrence of FP at two-loop in $\lambda_2(M_Z) - \alpha$ plane for different values of $\beta$ and $\gamma$. Initial value of Higgs-self quartic coupling $\lambda_1(M_Z) = 0.129$ for all cases. The values of $(\beta, \gamma)$ for the cases in (a)–(c), the $\beta$ value is 0.5 and $\gamma$ values are 0.1, 0.5, 1.0 respectively. For (d)–(f), the $\beta$ value is 1.0 and $\gamma$ values are 0.1, 0.5, 1.0 respectively. For (g)–(i), the $\beta$ value is 1.5 and $\gamma$ values are 0.1, 0.5, 1.0 respectively."

From Fig. 16 it is clear that there exist FP behaviors in the 2-loop evolution of $\lambda_1$ for certain initial values of $\lambda_2$, $\lambda_3$, $\lambda_4$, and $\lambda_5$. Figure 16(a) shows that for the choice of $\lambda_2 = 0.4, \alpha = 2.0, \beta = 1.0, \gamma = 0.1$, we see LP at one-loop at $10^7$ GeV and FP at the two-loop level at $10^{10}$ GeV, which is similar to $\phi^4$-theory, ISM and ITM. For an enhancement of $\lambda_2 = 0.8$, in Fig. 16(b), the LP and FP scales reduce to $10^4, 10^7$ GeV, respectively. At two-loop with $10^{18}$ GeV, we observe a jump and further FP behavior like ITM. In Fig. 16(c) as we increase the $\gamma = 1.0$, the LP and FP both disappear establishing $\gamma$ as a spoiler, which is the artefact of coupling involving $\lambda_5$. i.e., term with residual phase. As we sufficiently increase $\alpha = 3.5$ in Fig. 16(d), we get back both LP and FP, manifesting the $\alpha$ as an enhancer. This will be analyzed in detail in the following paragraphs.

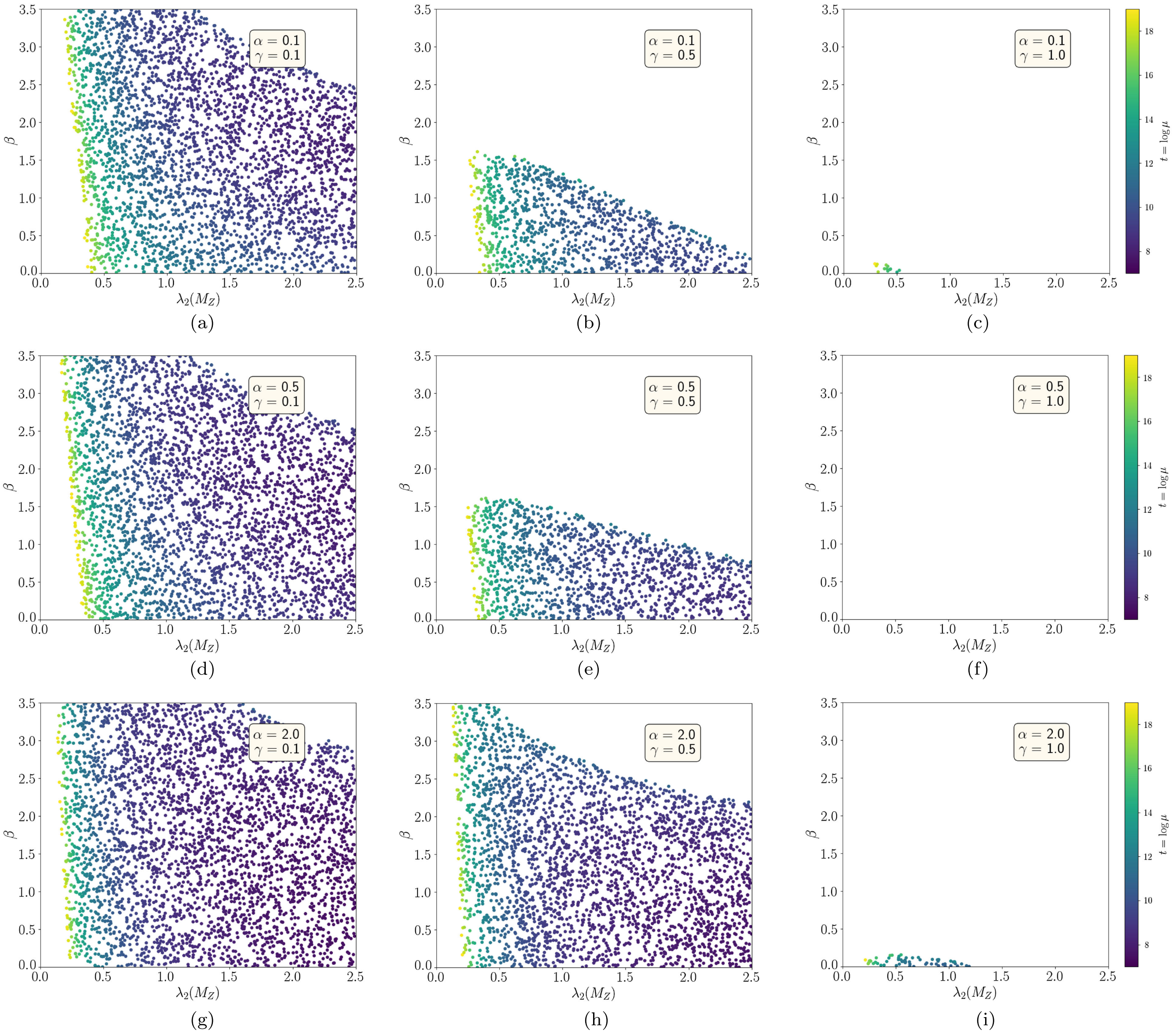

FIG. 19. Occurrence of FP at two-loop in $\lambda_2(M_Z) - \beta$ plane for different values of $\alpha$ and $\gamma$. Initial value of Higgs-self quartic coupling $\lambda_1(M_Z) = 0.129$ for all cases. The values of $(\alpha, \gamma)$ for the cases in (a)–(c), the $\alpha$ value is 0.1 and $\gamma$ values are 0.1, 0.5, 1.0 respectively. For (d)–(f), the $\alpha$ value is 0.5 and $\gamma$ values are 0.1, 0.5, 1.0 respectively. For (g)–(i), the $\alpha$ value is 2.0 and $\gamma$ values are 0.1, 0.5, 1.0 respectively.

In Fig. 17 we show the evolution of $\lambda_2$ and $\lambda_3$, similar to ISM and ITM, for $\alpha = 2.0$, $\beta = 1.0$, and $\gamma = 0.1$. The Fixed Points are observed for both $\lambda_2$ and $\lambda_3$ at the two-loop level around $10^{10}$ GeV. It is worth mentioning that, even though the FP value of $\lambda_2$, which is the self-coupling of the inert scalar, stays within the limit of $4\pi$, the FP value for Higgs portal coupling $\lambda_3$, crosses $4\pi$, which is similar to the case of ISM and ITM.

To study the role of the relative strengths of $\alpha$, $\beta$, and $\gamma$ we scatter the parameter points in the 2-D plane, where we vary one of the parameters with $\lambda_2$, and the scale $t = \log \mu$ is shown in the color code from blue (7) to yellow (19) while keeping the other two fixed in certain values as we look for the possible Fixed Points that start at certain renormalization scale $\mu$ at two-loop.

In Fig. 18, we present the variation of $\alpha$ vs $\mu$, where we parametrized some fixed values of $\beta, \gamma$. In Figs. 18(a)–18(c), we vary $\alpha$ with respect to $\lambda_2(M_Z)$ while taking $\gamma = 0.1, 0.5$, and 1.0, respectively for $\beta = 0.5$. There are two things that one can notice here; first, greater values of $\lambda_2(M_Z)$ correspond to the appearance of FP at lower scale $t = \log \mu$, and second, as we increase the value of $\gamma$ from (a) to (c), the number of points in the parameter space having Fixed Points, decreases. Thus $\gamma$ acts as a spoiler for such an FP. As we look for relatively higher values of $\beta = 1.0$ in Figs. 18(d)–18(f), the occurrence of FPs reduces further as compared to Figs. 18(a)–18(c), especially for the lower $\alpha$ values. Thus $\beta$ acts like a moderate spoiler. Further increment of $\beta$ to $\beta = 1.5$, lessen the occurrence of FPs and for $\beta = 1.5, \gamma = 1.0$ in Fig. 18(i) has no available FPs. For higher values of $\beta$, $\gamma$ similar observations are found.

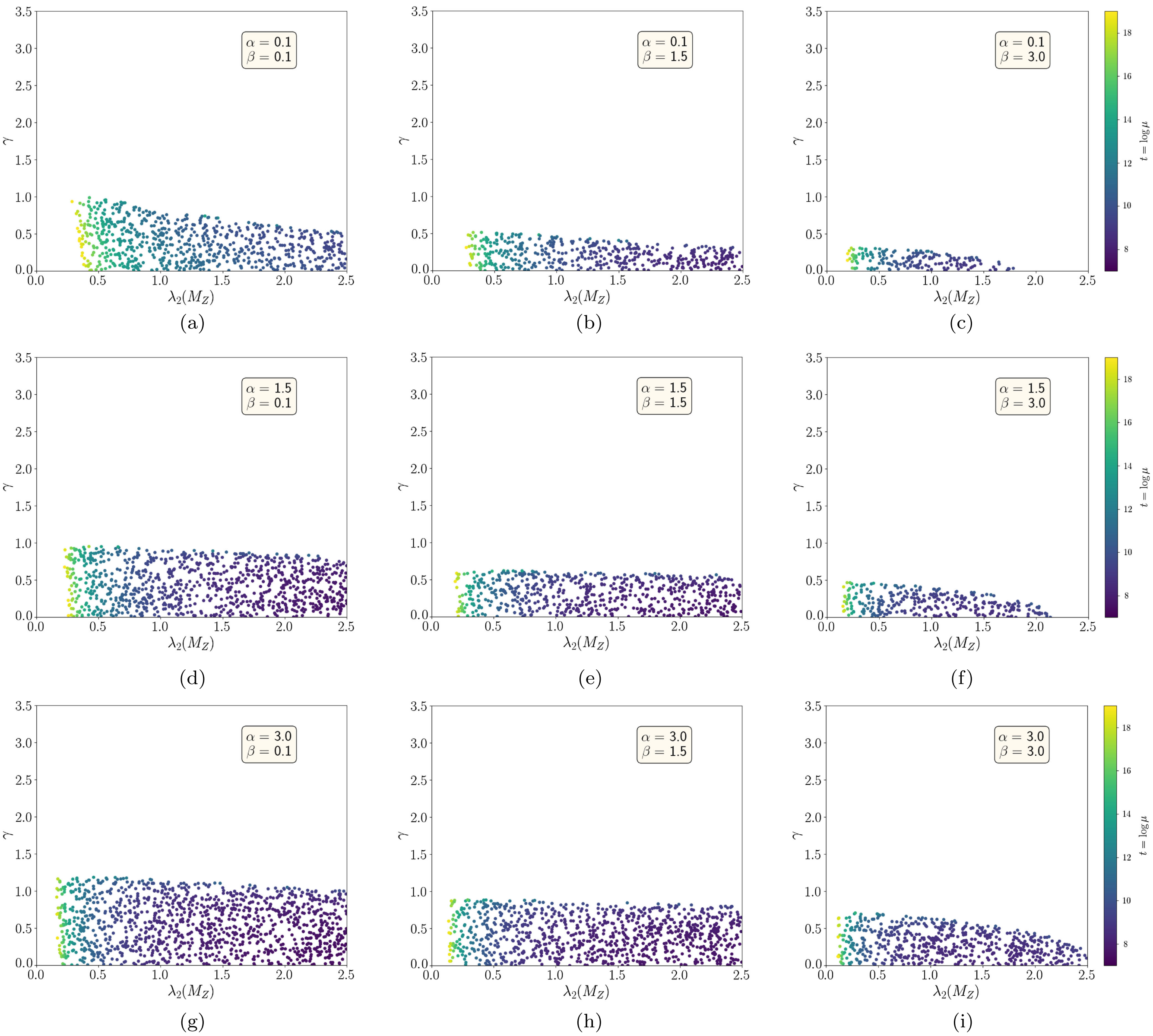

FIG. 20. Occurrence of FP at two-loop in $\lambda_2(M_Z) - \gamma$ plane for different values of $\alpha$ and $\beta$. Initial value of Higgs-self quartic coupling $\lambda_1(M_Z) = 0.129$ for all cases. The values of $(\alpha, \beta)$ for the cases in (a)–(c), the $\alpha$ value is 0.1 and $\beta$ values are 0.1, 1.5, 3.0 respectively. For (d)–(f), the $\alpha$ value is 1.5 and $\beta$ values are 0.1, 1.5, 3.0 respectively. For (g)–(i), the $\alpha$ value is 3.0 and $\beta$ values are 0.1, 1.5, 3.0 respectively.

In Figure 19 we vary $\beta$ verse $\lambda_2(M_Z)$ for different values of $\alpha$, $\gamma$. In Figs. 19(a)–19(c), where we enhance $\gamma$ from 0.1 to 1.0, keeping $\alpha = 0.1$, two-loop FP occurrences reduce successively from Figs. 19(a)–19(c) as $\gamma$ spoils heavily. For a given $\gamma$ value, an increase in $\alpha$ increases the FP occurrences for relatively smaller values of $\gamma$, which can be seen from Figs. 19(d), 19(e). However, for a larger value of $\gamma = 1.0$, the increase of $\alpha = 0.5$ is not sufficient, and Fig. 19(c) does not have any FP. Nevertheless, if we increase $\alpha$ to an adequate value of 2.0, FP reappears for all the three values of $\gamma$ as shown in Figs. 19(g)–19(i).

Finally, in Fig. 20, we check the occurrence of two-loop FPs in the $\lambda_2(M_Z) - \gamma$ plane for different values of $\alpha$ and $\beta$. In Figs. 20(a)–20(c), we show the Fixed Points for $\beta = 0.1$, 1.5, 3.0 for lower $\alpha = 0.1$. It is evident that the available FPs are getting lesser in numbers as we increase $\beta$ values. As we increase the $\alpha$ values column-wise in Figs. 20(d)–20(f) and 20(g)–20(i) for $\alpha = 1.5$, 3.0, the available FPs increase. This again establishes the behavior of $\alpha$ as an enhancer and $\beta$ as a moderate spoiler. $\gamma$ being the heavy spoiler, we do not see any Fixed Point for $\gamma \gtrsim 1$.

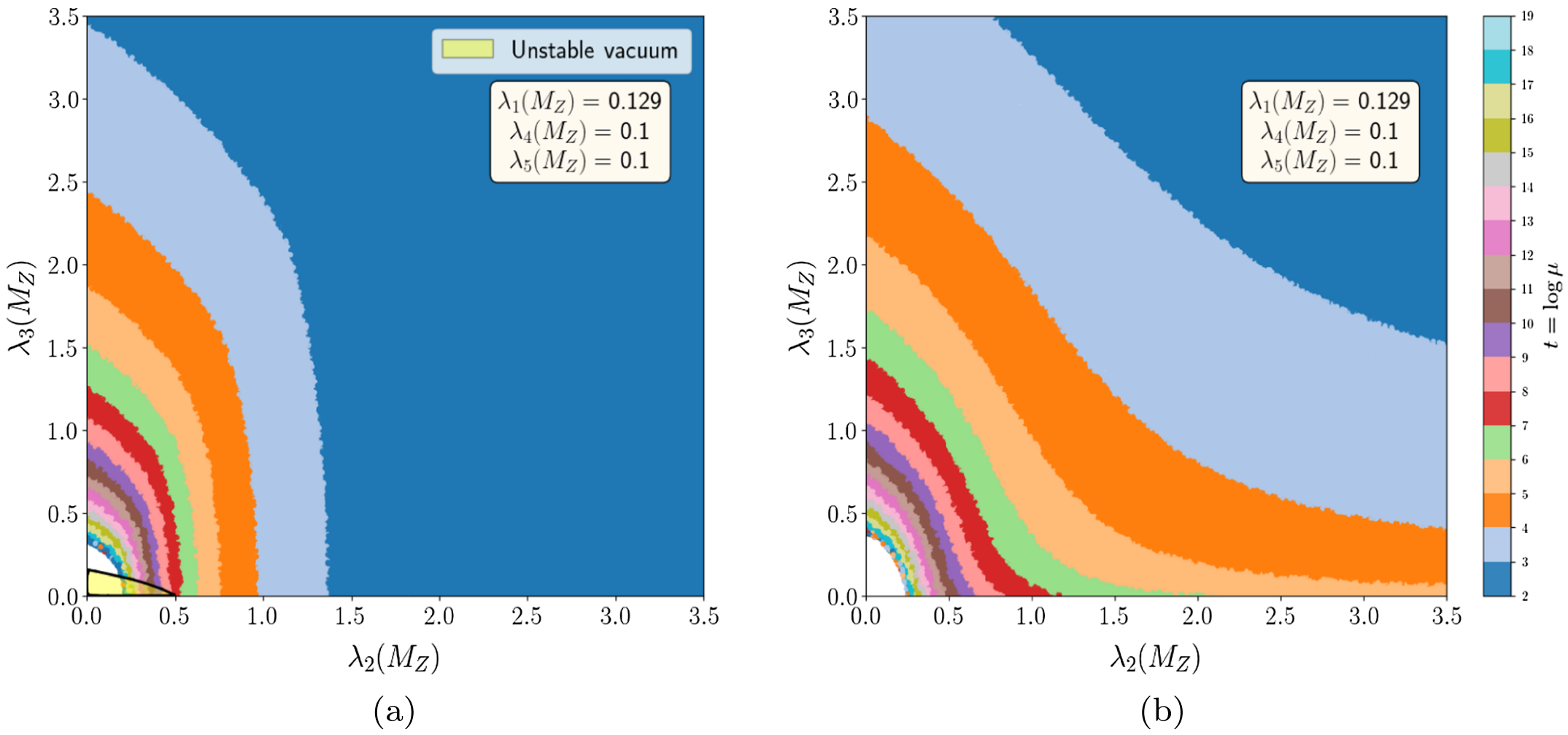

FIG. 21. Perturbativity limit of RGE evolution of Inert Doublet Model at one-loop level in (a) and two-loop level in (b). Initial value of Higgs-self quartic coupling $\lambda_H(M_Z) = 0.129$ and initial values of $\lambda_4(M_Z) = \lambda_5(M_Z) = 0.1$ for both cases.

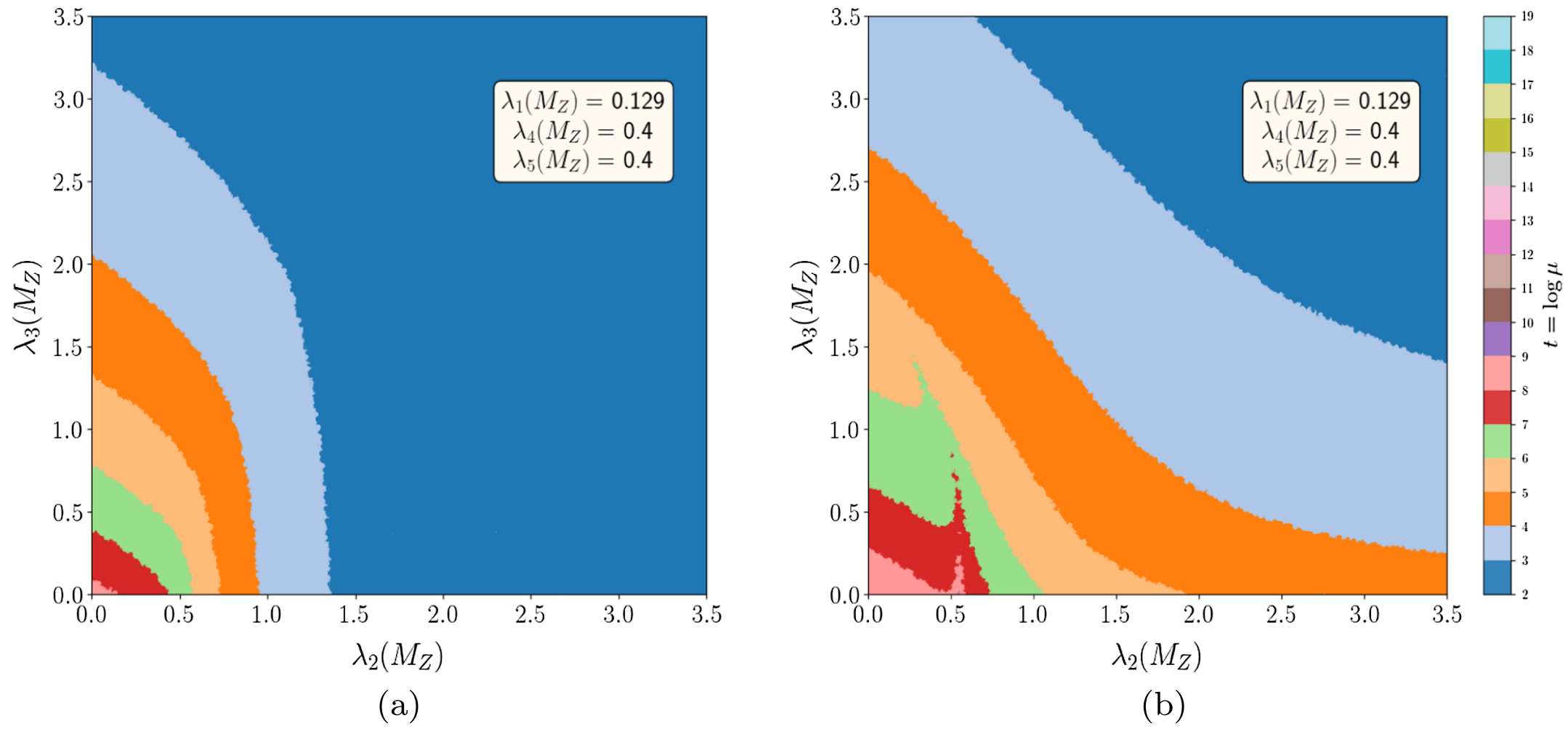

FIG. 22. Perturbativity limit of RGE evolution of Inert Doublet Model at one-loop level in (a) and two-loop level in (b). Initial value of Higgs-self quartic coupling $\lambda_H(M_Z) = 0.129$ and initial values of $\lambda_4(M_Z) = \lambda_5(M_Z) = 0.4$ for both cases.

### D. IDM perturbativity limits

In this subsection we put the perturbativity limits of $4\pi$, i.e., $\lambda_{1,2,3,4,5}(\mu) < 4\pi$ at one- and two-loop level, along with the other constraints given in Eq. (3.2), in $\lambda_2(M_Z) - \lambda_3(M_Z)$ plane for different scales with different combinations of $\lambda_4$, $\lambda_5$. Here $\lambda_2$ is the new physics self-coupling and $\lambda_3$ is the Higgs portal coupling just like in the cases of ISM and ITM. Additionally, we have mixing terms of $\lambda_4$, $\lambda_5$ and for the first case as described in Fig. 21 we consider these terms very small; $\lambda_4 = \lambda_5 = 0.1$ ensuring minimal interference. The color code for the perturbativity scale is the same as before. We notice that $\lambda_2 \lesssim 0.25$ and $\lambda_3 \lesssim 0.3$ ensures Planck scale perturbativity. For larger couplings at one-loop levels, $\lambda_2 \gtrsim 1.4$, $\lambda_3 \gtrsim 1.5$ the perturbativity scale boils down to $10^{2-3}$ GeV. For higher values of $\lambda_3$, the perturbative limits curve down to $10^{2-3}$ for even smaller values of $\lambda_2$. Contrary to the previous two cases of ISM and ITM, in the case of IDM, the two-loop perturbativity limits change their shapes for higher values of the couplings. For two-loop, higher values of $\lambda_2$, $\lambda_3$ are allowed at relatively larger perturbativity scales as in the case of ISM and ITM.

We can see in Fig. 21(a) the black-contoured yellow-shaded region which corresponds to the parameter region that fails to satisfy the bounded-from-below constraints that are necessary for a stable vacuum, shown in Eq. (4.15). The region has a maximum possible value of $\lambda_2(M_Z) \approx 0.5$ and a maximum possible value of $\lambda_3(M_Z) \approx 0.2$. However, the vacuum becomes stable for the entire parameter region in Fig. 21(b) at the two-loop level.

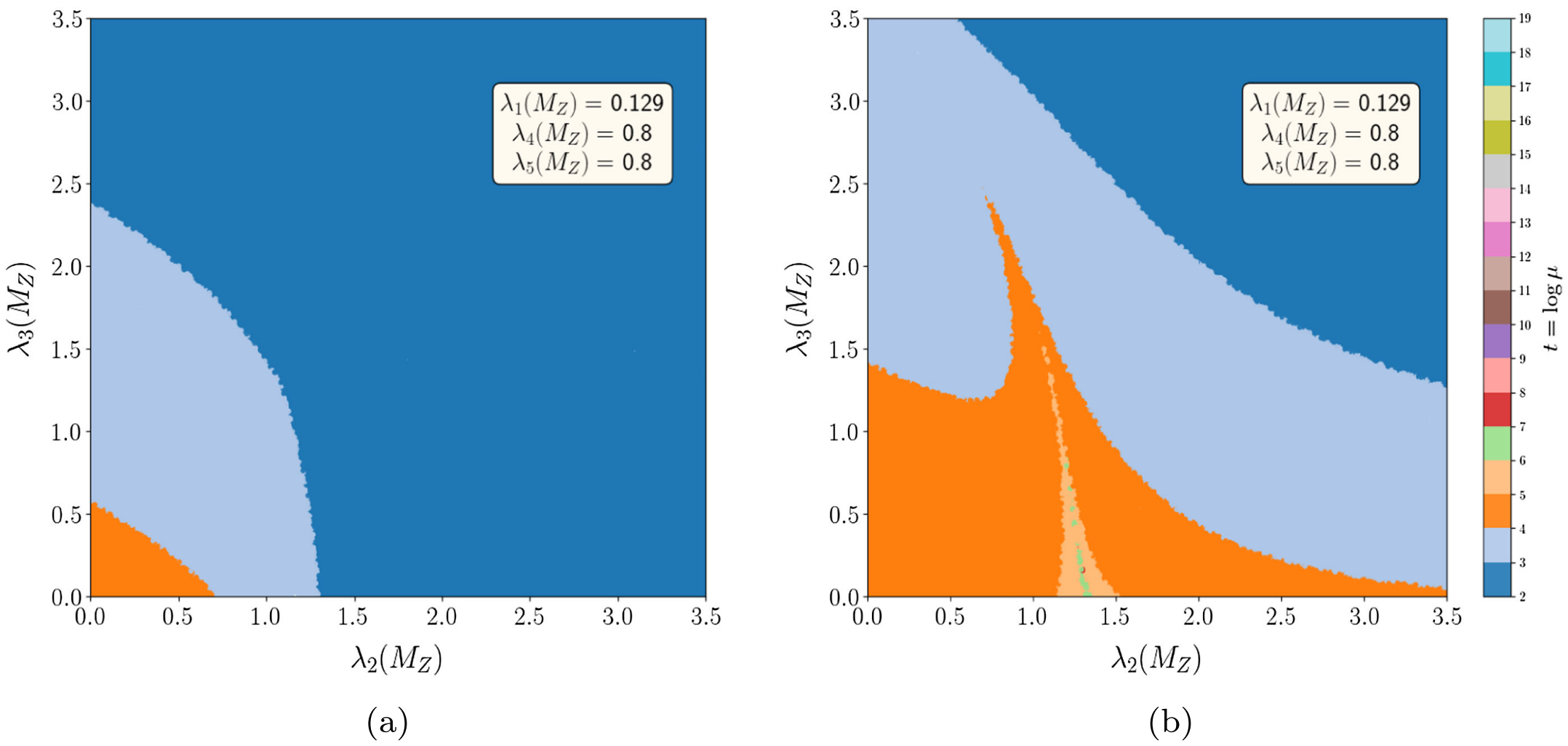

FIG. 23. Perturbativity limit of RGE evolution of Inert Doublet Model at one-loop level in (a) and two-loop level in (b). Initial value of Higgs-self quartic coupling $\lambda_H(M_Z) = 0.129$ and initial values of $\lambda_4(M_Z) = \lambda_5(M_Z) = 0.8$ for both cases.

Figure 22 describes the situation where the interfering couplings are relatively larger, i.e., $\lambda_4 = \lambda_5 = 0.4$. The interesting thing to observe is that there is no Planck scale perturbativity possible both at one- and two-loop levels. The maximum perturbative scale that can be achieved is $10^{8–9}$ GeV for lower values of the couplings and the minimum scale that is possible for higher coupling values is $10^{2–3}$ GeV. The two-loop limits are a little relaxed, like before, allowing larger values of the couplings. We also observe a kinklike nature near $\lambda_2 \sim 0.6$ at the two-loop level owing to the change of sign in the $\beta$-function. For both cases of one- and two-loop, the vacuum stability conditions Eq. (4.15) are satisfied for all the parameter regions in Fig. 22.

In Fig. 23 we enhance $\lambda_4$, $\lambda_5$ to 0.8. The maximum perturbativity limit comes down to $10^{4–5}$ GeV for lower couplings, while keeping the minimum limit to $10^{2–3}$ GeV for higher values of the couplings. Two-loop limits are more relaxed than one-loop, while the shapes are similar to the previous cases of $\lambda_4 = \lambda_5 = 0.4$. The kink regions, of course, expose the higher perturbativity limit, i.e. $10^{5–7}$ GeV. Similar to the previous case, for both one- and two-loop, the vacuum stability conditions Eq. (4.15) are satisfied by all points in the parameter space.

Figure 24 describes the case for $\lambda_4 = 0.4$, $\lambda_5 = 0.8$. In comparison to the previous case, as depicted in Fig. 23, lower couplings are now allowed for a higher perturbativity scale while the higher values of the coupling regions behave similarly. For an opposite choice i.e., $\lambda_4 = 0.8$, $\lambda_5 = 0.4$ as shown in Fig. 25, now lower values of $\lambda_2$ are allowed for relatively higher perturbativity scales. Similarly, the two-loop kink also shifts toward the left side in the $\lambda_2$ axis.

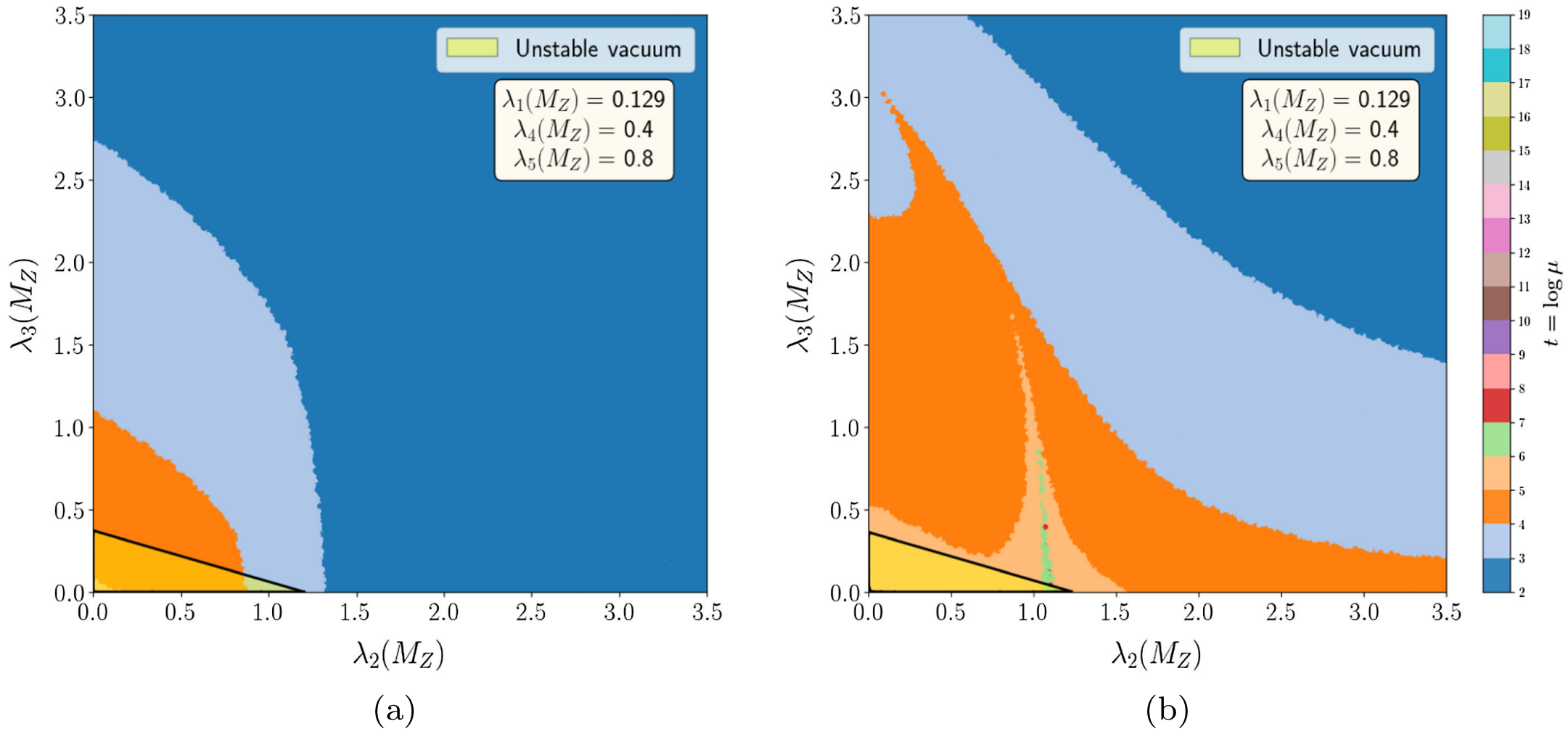

FIG. 24. Perturbativity limit of RGE evolution of Inert Doublet Model at one-loop level in (a) and two-loop level in (b). Initial value of Higgs-self quartic coupling $\lambda_H(M_Z) = 0.129$, initial values of $\lambda_4(M_Z) = 0.4$ and initial value of $\lambda_5(M_Z) = 0.8$ in both cases.

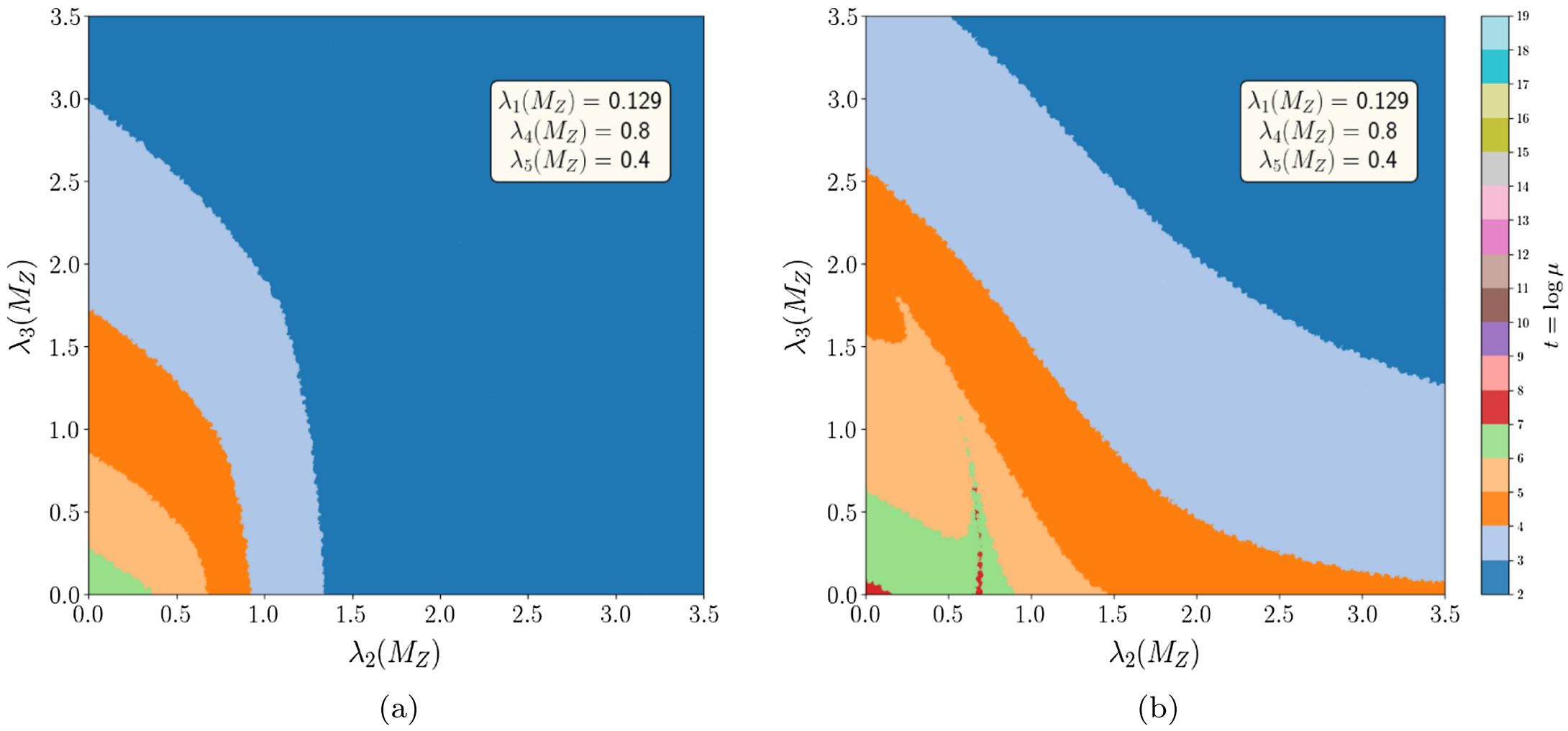

FIG. 25. Perturbativity limit of RGE evolution of Inert Doublet Model at one-loop level in (a) and two-loop level in (b). Initial value of Higgs-self quartic coupling $\lambda_H(M_Z) = 0.129$, initial values of $\lambda_4(M_Z) = 0.8$ and initial value of $\lambda_5(M_Z) = 0.4$ in both cases.

For $\lambda_4 = 0.4$, $\lambda_5 = 0.8$, in Figs. 24(a) and 24(b), we can see that there are points within the limits $\lambda_2 \leq 1.2$ and $\lambda_3 \leq 0.4$, for which the vacuum is unstable, as shown in the black-contoured yellow-shaded region. However, for the opposite choice $\lambda_4 = 0.8$, $\lambda_5 = 0.4$, in both the one- and two-loop plots, the vacuum is stable for all points as shown in Fig. 25.

## V. UNCERTAINTIES

The behavior of the running couplings as expected depends on the initial values of the coupling constants, viz. $\lambda_H(M_Z), Y_t(M_Z)$. The initial values of these two couplings can be recast from the uncertainties of the mass values of the Higgs boson and the top quark. The most recent Higgs mass measurements from combined Run I and Run II data at the ATLAS detector of LHC gives $1\sigma$ uncertainty as $m_H = 125.22 \pm 0.11(\text{stat}) \pm 0.09(\text{syst})$ GeV [80]. Similarly the global average value from both Tevatron and LHC of the top quark mass is given by $172.57 \pm 0.29$ GeV [81]. Considering the uncertainties of the Higgs mass, which in turn changes the $\lambda_H(M_Z)$ will affect the evolution of the $\lambda_H(\mu)$ and change the value for given scale $\mu$ for a fixed top quark mass. Defining the uncertainty as $\eta(\mu) = \frac{\lambda_H^{\pm}(\mu) - \lambda_H(\mu)}{\lambda_H(\mu)} = \frac{\Delta\lambda_H(\mu)}{\lambda_H(\mu)}$, where $\lambda_H^{\pm}(\mu)$ is the value of $\lambda_H$ corresponding to the initial values with the uncertainties. For $\lambda_H(\mu)$ values before achieving the Fixed Point, the uncertainty over both Higgs mass and top mass can vary from 3% to 40%, depending on the scale $\mu$. For example in case of ISM, closer to the $\lambda_H(\mu) = 0$, such uncertainty goes to 40%, however for more positive values of $\lambda_H(\mu)$ it can be as low as 3% for $\mu = 10^6$ GeV. However, once it achieves the Fixed Point, the values of the coupling becomes independent on the initial values of the running coupling, higher initial values initiate the FP earlier than the lower initial values. Thus it is the scale at which the coupling constant achieves FP changes but not the value of Fixed Points. Nevertheless, we checked numerically and for all three scenarios: ISM, ITM, and IDM such uncertainties stay within $(10^{-4}\text{–}10^{-5})\%$ for Higgs mass variation and $(10^{-3}\text{–}10^{-5})\%$ for the top quark mass variation.

## VI. PHENOMENOLOGICAL RELEVANCE

The perturbativity bounds at one- and two-loops for the three different inert models are mass independent theoretical bounds. However, most of the bounds, i.e., from electroweak precision data (EWPD) [16,33,34,39–41,81], FOPT [34,47,53,54,59,82], collider searches [16,24,33,36,41–43,81,83], dark matter relic and direct dark matter detection [14,16,17,25–27,33,37,40,43,79,84] are dependent of masses along with the portal or other corresponding couplings. Thus for the phenomenological relevant points we need to invoke masses along with the bounds on the scalar quartic couplings from the perturbativity. Let us consider the S, T, U parameters, where the singlet does not contribute, and only bounds come to IDM [24,33,34] and ITM [16,39–41]. From Fig. 1 of [33] we can see that for IDM, $\lambda_3 \sim 1$ for $M_{H^\pm} \sim 200$ GeV is allowed by the EWPD and $h \to \gamma\gamma$ decay width. Such low mass points may have potential FOPT, however, they are severely constrained by the most recent direct dark matter measurement LUX-ZEPLIN [84], which demands $\lambda_{345} = \lambda_3 + \lambda_4 - 2\lambda_5 \lesssim 0.01$. For example $m_{A^0} = 201$, $m_{H^0} = 325$, $m_{H^\pm} = 265$ GeV and the coupling choice of $\lambda_3 = 1$, $\lambda_4 = 0.0945$, $\lambda_5 = 0.542$ range a typical charged Higgs boson or heavy Higgs bosons pair or associated production cross section at 14 TeV LHC are around $(\mathcal{O})(6\text{–}18)$ fb, which are mediated by the gauge bosons, and similar studies can be found in the literature [32,85–87]. However, such productions do not depends on the portal couplings and one has to look for gluon mediation to SM Higgs followed by the discalar production via the Higgs portal couplings. A charged Higgs pair via gluon fusion at the LHC with 14 TeV center of mass energy is around 0.42 fb. Thus, we need very high luminosity and better algorithm to probe $\lambda_3$. Similarly, we can look for other Higgs pair from the gluon fusion that would enable us to probe the other portal couplings. Nonetheless, such cross sections depend on the combinations of the portal couplings a details study is required for this, which is beyond the scope of this article. For the above choice of portal couplings, a choice of $\lambda_2(M_Z) = 0.8$, gives rise to a Fixed Point around $10^8$ GeV. For a charged Higgs mass of 1 TeV, even though $\lambda_3(M_Z) \sim 4\pi$ is allowed by $h \to \gamma\gamma$ [33], the criteria of FP along with LUX-ZEPLIN bound make it difficult to achieve a FP. One has to carefully check the combination of $\alpha, \beta, \gamma$ and LUX-ZEPLIN bound to look for the possible FPs, and a higher $\lambda_3 \gtrsim 3$ is highly disfavored as can be seen from Fig. 18.

On the other hand, the most recent search for ITM dark matter reveals that there is no allowed point which satisfies the correct observed DM relic [82] and the maximum value of an underabundant triplet dark matter mass is $\lesssim$1.6 TeV. The LUX-ZEPLIN bound allows $\lambda_{HT} < 0.02, 0.2$ for $m_T = 300$, 1600 GeV. respectively. For ITM with the maximum allowed values of $\lambda_{HT}$ and for a higher mass ($\gtrsim$1 TeV), the normal gauge boson mediated processes along with the discalar production via gluon fusion are very suppressed. Thus one can look for the vector boson fusion (VBF) modes at the LHC and the muon collider [41,42,88–90]. For the points with $\lambda_{HT} = 0.02$, 0.2 and for a $\lambda_T = 0.8$, a FP scale can be achieved around $\mu = 10^{15}, 10^{12}$ GeV, respectively.

Finally, for ISM a closer study reveals that with new LUX-ZEPLIN bound [84], except for the $s$-channel SM Higgs resonance, i.e., $m_S \sim \frac{m_h}{2}$, no other points are allowed. For $m_S \sim \frac{m_h}{2}$, a maximum value of $\lambda_{HS} \lesssim 0.03$ is still allowed. A discalar production cross section via gluon fusion is around 1 fb at the LHC with 14 TeV centre of mass energy. Probing this complete missing final state one has to look for monophoton or mono-jet plus missing energy final state [91–94]. For the same portal coupling with a mass of 1 TeV, such cross section however drops drastically to $10^{-8}$ fb. Thus we have to look for it at future circular collider (FCC-hh). A FP point can be achieved for this choice of portal coupling with $\lambda_S = 1.0$ at the scale $\mu \sim 10^{14}$ GeV. All in all, satisfying everything for a single inter scenario can be challenging and a multiscalar extension is more feasible [95].

## VII. DISCUSSIONS AND CONCLUSION

In this article, we investigate the RG evolution of the scalar quartic couplings for the inert models, namely inert singlet, inert doublet, and Inert Triplet Models. An interesting feature of having a Landau pole at one-loop and a Fixed Point at two-loops for ISM and ITM, which is very similar to the $\phi^4$ theory, is noticed. In the $\phi^4$-theory, the existence of Landau poles for odd-loops and Fixed Points for even loops and their reliability are studied till seven-loop beta-functions [65,66]. However, for the Standard Model, no such observation is found till the two-loop level as the initial values of parameters are fixed by the experimental data.

Extending the SM with other scalars: namely ISM, ITM, and IDM, we could invoke more parameters and see the possibility of FPs at the two-loop level in the scalar sector, especially for the Higgs quartic coupling. We saw that the evolution of SM-like Higgs quartic coupling attains a two-loop FP for certain initial values of parameters. It is mainly governed by the relative strength of portal coupling with the strength of self-coupling of the inert scalar. We observe that in the case of inert singlet and triplet, for small initial values of self-coupling ($\lambda_S$ or $\lambda_T$), the occurrence of such FPs is promoted by the increase in initial values of the portal couplings ($\lambda_{HS}$ or $\lambda_{HT}$). For higher initial values of $\lambda_S$ or $\lambda_T$, there is a limit at which portal couplings can promote FPs. Unlike these two cases, the inert doublet has three different portal couplings ($\lambda_3$, $\lambda_4$, and $\lambda_5$) and the latter two behave differently than $\lambda_3$. The portal coupling ($\lambda_3$) that resembles the singlet/ triplet case, the coefficient of the term that only depends on the square of absolute values of the component fields, promotes the occurrence of such FPs, whereas portal couplings $\lambda_4$ and $\lambda_5$, which correspond to the terms that have residual phases, suppress the occurrence of FPs. Thus, the appearance of FPs in IDM is not so obvious as $\lambda_4$ and $\lambda_5$ act as spoilers. Nevertheless, a sufficiently larger value of $\lambda_3$ relative to $\lambda_{4,5}$ can bring back the FPs. The residual existing phases from the $\lambda_4$, $\lambda_5$ terms are instrumental in spoiling the FPs at the two-loop level [66]. In this analysis, we inherited a basis of $\alpha$, $\beta$, $\gamma$, which are the relative strength of $\lambda_3$, $\lambda_4$, $\lambda_5$ with respect to $\lambda_2$, respectively. We found that while the $\alpha$ behaves like an enhancer of the FPs, $\beta$ and $\gamma$ behave like moderate and heavy spoilers of the FP. This will erudite us about the particle behavior involving the scalar quartic couplings, which will lead to a better understanding of FPs in quantum field theories and help in realizing how they manifest in low-energy interactions.

Considering the conditions for stable vacuum, we see that most of the regions with lower new physics couplings, i.e., $\lambda_{i\neq 1} \lesssim 1.0$ are ruled out, at least for ISM and ITM. However, the situation gets interesting for IDM, as for $\lambda_4 = \lambda_5 \gtrsim 0.4$, other couplings, i.e., $\lambda_{i=2,3}$ are also allowed even for smaller values, i.e., $\lesssim$1. A combination of lower $\lambda_4$ and greater $\lambda_5$, viz., $\lambda_4 = 0.4$, $\lambda_5 = 0.8$, the lower values of other couplings correspond to the unstable vacuum. Thus, perturbative limits are more relevant for higher self and portal couplings.

It is also observed that higher quartic couplings are bounded by relatively lower perturbativity scales,

restricting the theory to lower scales. Higher portal couplings are often favored for FOPT [83] and even for getting the correct or under-abundant dark matter relics [32]. In particular, a TeV scale dark matter prefers larger portal couplings [32,33,41,42]. For the scalars in the higher representations of $SU(2)$, to satisfy the current measurement of Higgs to diphoton rate prefers relatively larger portal couplings that are still allowed [24,33,41–43]. If we consider this perturbative unitarity bound on the quartic couplings $\gtrsim 1$, the theory gets nonperturbative at the scale $10^{2-3}$ GeV. In those cases, a multi-TeV dark matter with corresponding large quartic couplings, which are also required to get FOPT, would be questionable while considering perturbativity. Considering current LUX-ZEPLIN [84], such points are heavily constrained, especially for ISM and ITM, and very fine tuned regions can be allowed for IDM as discussed in this article. Thus a single inert model is unfeasible to explain all the things together, and a multiscalar scenario can be a good option [95]. Current collider bounds on the scalar quartic couplings are very weak from the di-Higgs production [96–102], which makes these theoretical estimations more relevant. A future muon collider and FCC-hh can probe some of these portal couplings.

The asymptotic behaviors of the scalar quartic couplings for all three inert models are similar as they increase with the energy scale $\mu$ at two-loop till they attain a Fixed Point for a judicial choice of the initial values. If the coupling constant at the FP stay below $4\pi$ that makes the theory perturbatively predictive, otherwise it becomes unpredictive. The generic trends for all three scenarios are similar. Mostly the SM Higgs self-coupling, $\lambda_H$ and the new physics self-coupling $\lambda_{S/T/2}$ stays below $4\pi$, while the Higgs portal couplings tend to break it by crossing the limit of $4\pi$. For example in ISM, the Fixed Point for $\lambda_H$ appears around $10^{10}$ GeV for a choice of $\lambda_{HS}(M_Z) = \lambda_S(M_Z) = 0.8$, which remains in the finite region of $\lambda_H < 4\pi$, thus any scattering cross section or decay involving $\lambda_H$ remains predictive. However, any such calculation involving the Higgs portal couplings will violate the perturbative unitarity constraints. On the other hand, for ITM, a similar choice of the initial condition, attains the FP around $10^{11}$ GeV and $\lambda_H$ remains $< 4\pi$ till $10^{18}$ GeV, although, soon after that it attains another FP, which lies above $4\pi$. Thus the asymptotic behavior becomes unpredictable or fluctuating for very high energy scales, and this behavior becomes prominent at relatively lower scale as we increase the initial values of the couplings. An interesting behavior is observed for both $\lambda_{HT}, \lambda_T$, as around $10^{18}$ GeV both the couplings come down below $4\pi$ leaving their FP behavior. For IDM a lower value of $\lambda_2(M_Z), \beta, \gamma$ and a higher $\alpha$, the SM Higgs coupling $\lambda_H$ attains a FP around $10^{10}$ GeV, which remains perturbative till the Planck scale. Although, for a choice of higher $\lambda_2(M_Z) = 0.8$, we discover another FP around $10^{18}$ GeV, where $\lambda_H$ crosses $4\pi$ very similar to ITM. The portal and new physics self-couplings behavior remain similar to ISM.

## ACKNOWLEDGMENTS

P. B. wants to acknowledge Saumen Datta, Tuhin Roy, Namit Mahajan, Anirban Kundu, Anirban Karan for the valuable discussion during the Flavour Physics Week at IITGn. P. B. also wants thank Luigi Delle Rose for his comments and suggestions. P. M. P. R. wants to thank Chandrima Sen and Disha Ghosh for their help in different stages of the project. P. B. and P. M. P. R. wants to mention their special gratitude to Snehashis Parashar for his valuable inputs in the Phenomenology discussion. P. B. and P. M. P. R. want to thank Myriam Mondragón for useful inputs during PPC 2024. P. M. P. R. also acknowledges Ministry of Education, India funding for the research work.

## DATA AVAILABILITY

The data that support the findings of this article are not publicly available. The data are available from the authors upon reasonable request.

## APPENDIX: β-FUNCTIONS OF THE SM AND THE INERT MODELS AT TWO-LOOP LEVEL

### 1. Standard Model

$$\beta_{g_1}^{\mathrm{SM}} = \frac{1}{(4\pi)^2}\left(\frac{41}{10}g_1^3\right) + \frac{1}{(4\pi)^4}\left(\frac{199}{200}g_1^5 + \frac{27}{40}g_1^3 g_2^2 + \frac{11}{5}g_1^3 g_3^2 - \frac{17}{40}g_1^3 Y_t^2\right) \tag{A1}$$

$$\beta_{g_2}^{\mathrm{SM}} = \frac{1}{(4\pi)^4}\left(\frac{9}{40}g_1^2 g_2^3 + \frac{35}{24}g_2^5 + 3g_2^3 g_3^2 - \frac{3}{8}g_2^3 Y_t^2\right) + \frac{1}{(4\pi)^2}\left(\frac{19}{6}g_2^3\right) \tag{A2}$$

$$\beta_{g_3}^{\mathrm{SM}} = \frac{1}{(4\pi)^4}\left(\frac{11}{40}g_1^2 g_3^3 + \frac{9}{8}g_2^2 g_3^3 - \frac{13}{2}g_3^5 - \frac{1}{2}g_3^3 Y_t^2\right) + \frac{1}{(4\pi)^2}(7g_3^3) \tag{A3}$$

$$\beta_{Y_t}^{\rm SM} = \frac{1}{(4\pi)^2}\left(\frac{9}{32}Y_t^3 + \frac{17}{320}g_1^2 Y_t - \frac{9}{64}g_2^2 Y_t - \frac{1}{2}g_3^2 Y_t\right) + \frac{1}{(4\pi)^4}\left(\frac{1187}{153600}g_1^4 Y_t + \frac{19}{3840}g_1^2 g_3^2 Y_t + \frac{9}{256}g_2^2 g_3^2 Y_t + \frac{3}{128}\lambda^2 Y_t + \frac{3}{64}Y_t^5 + \frac{3}{64}\lambda Y_t^3 - \frac{9}{5120}g_1^2 g_2^2 Y_t - \frac{23}{1024}g_2^4 Y_t - \frac{27}{64}g_3^4 Y_t - \frac{393}{20480}g_1^2 Y_t^3 - \frac{225}{4096}g_2^2 Y_t^3 - \frac{9}{64}g_3^2 Y_t^3 - \frac{9}{32}Y_t^3\right)$$

$$\beta_{\lambda}^{\rm SM} = \frac{1}{(4\pi)^2}\left(\frac{27}{800}g_1^4 + \frac{9}{80}g_1^2 g_2^2 + \frac{9}{32}g_2^4 - \frac{9}{20}g_1^2\lambda + 6\lambda^2 + 3\lambda Y_t^2 - \frac{9}{4}g_2^2\lambda - \frac{3}{2}Y_t^4\right) + \frac{1}{(4\pi)^4}\left(\frac{305}{16}g_2^6 + \frac{1887}{200}g_1^4\lambda + \frac{117}{20}g_1^2 g_2^2\lambda + \frac{108}{5}g_1^2\lambda^2 + 108 g_2^2\lambda^2 + \frac{63}{10}g_1^2 g_2^2 Y_t^2 + 30 Y_t^6 + \frac{17}{2}g_1^2\lambda Y_t^2 + \frac{45}{2}g_2^2\lambda Y_t^2 + 80 g_3^2\lambda Y_t^2 - \frac{3411}{2000}g_1^6 - \frac{1677}{400}g_1^4 g_2^2 - \frac{289}{80}g_1^2 g_2^4 - \frac{73}{8}g_2^4\lambda - 312\lambda^3 - \frac{171}{100}g_1^4 Y_t^2 - \frac{9}{4}g_2^4 Y_t^2 - 144\lambda^2 Y_t^2 - \frac{8}{5}g_1^2 Y_t^4 - 32 g_3^2 Y_t^4 - 3\lambda Y_t^4\right) \tag{A4}$$

## 2. Inert Singlet Model

$$\beta_{g_1}^{\rm IS} = \beta_{g_1}^{\rm SM}, \qquad \beta_{g_2}^{\rm IS} = \beta_{g_2}^{\rm SM}, \qquad \beta_{g_3}^{\rm IS} = \beta_{g_3}^{\rm SM}, \qquad \beta_{Y_t}^{\rm IS} = \beta_{Y_t}^{\rm SM} + \frac{1}{2}\lambda_{HS}^2 Y_t, \tag{A5}$$

$$\beta_{\lambda_H}^{\rm IS} = \beta_{\lambda_H}^{\rm SM} + \frac{1}{(4\pi)^2}\left(\frac{1}{16}\lambda_{HS}^2\right) - \frac{1}{(4\pi)^4}\left(\frac{5}{128}\lambda_H\lambda_{HS}^2 + \frac{1}{64}\lambda_{HS}^3\right).$$

$$\beta_{\lambda_{HS}}^{\rm IS} = \frac{1}{(4\pi)^2}\left(12\lambda_H\lambda_{HS} + 4\lambda_{HS}^2 + 8\lambda_{HS}\lambda_S + 6\lambda_{HS}Y_t^2 - \frac{9}{10}g_1^2\lambda_{HS} - \frac{9}{2}g_2^2\lambda_{HS}\right) + \frac{1}{(4\pi)^4}\left(\frac{1671}{400}g_1^4\lambda_{HS} + \frac{9}{8}g_1^2 g_2^2\lambda_{HS} + \frac{72}{5}g_1^2\lambda_H\lambda_{HS} + 72 g_2^2\lambda_H\lambda_{HS} + \frac{3}{5}g_1^2\lambda_{HS}^2 + 3 g_2^2\lambda_{HS}^2 + \frac{17}{4}g_1^2\lambda_{HS}Y_t^2 + \frac{45}{4}g_2^2\lambda_{HS}Y_t^2 + 40 g_3^2\lambda_{HS}Y_t^2 - 72\lambda_H\lambda_{HS}Y_t^2 - 12\lambda_{HS}^2 Y_t^2 - \frac{27}{2}\lambda_{HS}Y_t^4 - \frac{145}{16}g_2^4\lambda_{HS} - 60\lambda_H^2\lambda_{HS} - 72\lambda_H\lambda_{HS}^2 - 11\lambda_{HS}^3 - 48\lambda_{HS}^2\lambda_S - 40\lambda_{HS}\lambda_S^2\right)$$

$$\beta_{\lambda_S}^{\rm IS} = \frac{1}{(4\pi)^2}(2\lambda_{HS}^2 + 20\lambda_S^2) + \frac{1}{(4\pi)^4}\left(\frac{12}{5}g_1^2\lambda_{HS}^2 + 12 g_2^2\lambda_{HS}^2 - 8\lambda_{HS}^3 - 20\lambda_{HS}^2\lambda_S - 240\lambda_S^3 - 12\lambda_{HS}^2 Y_t^2\right) \tag{A6}$$

## 3. Inert Triplet Model

$$\beta_{g_1}^{\rm ITM} = \beta_{g_1}^{\rm SM}, \qquad \beta_{g_2}^{\rm ITM} = \beta_{g_2}^{\rm SM} + \frac{1}{(4\pi)^2}\left(\frac{1}{3}g_2^3\right) + \frac{1}{(4\pi)^4}\left(\frac{28}{3}g_2^5\right) \tag{A7}$$

$$\beta_{g_3}^{\rm ITM} = \beta_{g_3}^{\rm SM}, \qquad \beta_{Y_t}^{\rm ITM} = \beta_{Y_t}^{\rm SM} + \frac{1}{(4\pi)^4}\left(g_2^4 Y_t^2 + \frac{3}{4}\lambda_{HT}^2 Y_t^2\right) \tag{A8}$$

$$\beta_{\lambda_H}^{\rm ITM} = \beta_{\lambda_H}^{\rm SM} + \frac{1}{(4\pi)^2}\left(\frac{3}{2}\lambda_{HT}^2\right) + \frac{1}{(4\pi)^4}\left(\frac{11}{2}g_2^4\lambda_H + \frac{5}{4}g_2^4\lambda_{HT} + \frac{9}{2}g_2^2\lambda_{HT}^2 - 15\lambda_H\lambda_{HT}^2 - 2\lambda_{HT}^3 - \frac{7}{20}g_1^2 g_2^4 - \frac{7}{4}g_2^6\right) \tag{A9}$$

$$\beta_{\lambda_{HT}}^{\rm ITM} = \frac{1}{(4\pi)^4}\left(12\lambda_H\lambda_{HT} + 10\lambda_{HT}\lambda_T + 6\lambda_{HT}Y_t^2 - \frac{9}{10}g_1^2\lambda_{HT} - \frac{33}{2}g_2^2\lambda_{HT}\right)$$
$$+ \frac{1}{(4\pi)^4}\left(\frac{1671}{400}g_1^4\lambda_{HT} + \frac{9}{8}g_1^2g_2^2\lambda_{HT} + \frac{72}{5}g_1^2\lambda_H\lambda_{HT} + 72g_2^2\lambda_H\lambda_{HT} + \frac{5}{2}g_2^4\lambda_T\right.$$
$$+ 28g_2^2\lambda_{HT}\lambda_T + \frac{17}{4}g_1^2\lambda_{HT}Y_t^2 + \frac{45}{4}g_2^2\lambda_{HT}Y_t^2 + 40g_3^2\lambda_{HT}Y_t^2 - \frac{209}{6}g_2^4\lambda_{HT}$$
$$\left. - 60\lambda_H^2\lambda_{HT} - \frac{23}{2}\lambda_{HT}^3 - 4\lambda_{HT}^2\lambda_T - 26\lambda_{HT}\lambda_T^2 - 72\lambda_H\lambda_{HT}Y_t^2 - \frac{27}{2}\lambda_{HT}Y_t^4\right) \tag{A10}$$

$$\beta_{\lambda_T}^{\rm ITM} = \frac{1}{(4\pi)^4}\left(\frac{3}{8}g_2^4 + 2\lambda_{HT}^2 + 18\lambda_T^2 - 24g_2^2\lambda_T\right) + \frac{1}{(4\pi)^4}\left(\frac{12}{5}g_1^2\lambda_{HT}^2 + 12g_2^2\lambda_{HT}^2 + 54g_2^2\lambda_T^2\right.$$
$$\left. - \frac{55}{24}g_2^6 - \frac{653}{12}g_2^4\lambda_T - 12\lambda_{HT}^2\lambda_T - 156\lambda_T^3 - 12\lambda_{HT}^2Y_t^2\right) \tag{A11}$$

### 4. Inert Doublet Model

$$\beta_{g_1}^{\rm IDM} = \beta_{g_1}^{\rm SM} + \frac{1}{(4\pi)^2}\left(\frac{1}{10}g_1^3\right) + \frac{1}{(4\pi)^4}\left(\frac{9}{50}g_1^5 + \frac{9}{10}g_1^3g_2^2\right) \tag{A12}$$

$$\beta_{g_2}^{\rm IDM} = \beta_{g_2}^{\rm SM} + \frac{1}{(4\pi)^2}\left(\frac{1}{6}g_2^3\right) + \frac{1}{(4\pi)^4}\left(\frac{3}{10}g_1^2g_2^3 + \frac{13}{6}g_2^5\right) \tag{A13}$$

$$\beta_{g_3}^{\rm IDM} = \beta_{g_3}^{\rm SM} \tag{A14}$$

$$\beta_{Y_t}^{\rm IDM} = \beta_{Y_t}^{\rm SM} + \frac{1}{(4\pi)^4}\left(\frac{2}{15}g_1^4Y_t + \frac{1}{2}g_2^4Y_t + \lambda_3^2Y_t + \lambda_3\lambda_4Y_t + \lambda_4^2Y_t + 6\lambda_5^2Y_t\right) \tag{A15}$$

$$\beta_{\lambda_1}^{\rm IDM} = \beta_{\lambda_1}^{\rm SM} + \frac{1}{(4\pi)^2}(2\lambda_3^2 + 2\lambda_3\lambda_4 + \lambda_4^2 + 4\lambda_5^2) + \frac{1}{(4\pi)^4}\left(\frac{33}{100}g_1^4\lambda_1 + \frac{11}{4}g_2^4\lambda_1 + \frac{9}{10}g_1^4\lambda_3\right.$$
$$+ \frac{15}{2}g_2^4\lambda_3 + \frac{12}{5}g_1^2\lambda_3^2 + 12g_2^2\lambda_3^2 + \frac{9}{20}g_1^4\lambda_4 + \frac{3}{2}g_1^2g_2^2\lambda_4 + \frac{15}{4}g_2^4\lambda_4 + \frac{12}{5}g_1^2\lambda_3\lambda_4 + 12g_2^2\lambda_3\lambda_4$$
$$+ \frac{6}{5}g_1^2\lambda_4^2 + 3g_2^2\lambda_4^2 - \frac{12}{5}g_1^2\lambda_5^2 - \frac{63}{1000}g_1^6 - \frac{21}{200}g_1^4g_2^2 - \frac{7}{40}g_1^2g_2^4 - \frac{7}{8}g_2^6 - 20\lambda_1\lambda_3^2 - 8\lambda_3^3 - 20\lambda_1\lambda_3\lambda_4$$
$$\left. - 12\lambda_3^2\lambda_4 - 12\lambda_1\lambda_4^2 - 16\lambda_3\lambda_4^2 - 6\lambda_4^3 - 56\lambda_1\lambda_5^2 - 80\lambda_3\lambda_5^2 - 88\lambda_4\lambda_5^2\right) \tag{A16}$$

$$\beta_{\lambda_2}^{\rm IDM} = \frac{1}{(4\pi)^2}\left(\frac{27}{200}g_1^4 + \frac{9}{20}g_1^2g_2^2 + \frac{9}{8}g_2^4 + 24\lambda_2^2 + 2\lambda_3^2 + 2\lambda_3\lambda_4 + \lambda_4^2 + 4\lambda_5^2 - \frac{9}{5}g_1^2\lambda_2 - 9g_2^2\lambda_2\right)$$
$$+ \frac{1}{(4\pi)^4}\left(\frac{291}{16}g_2^6 + \frac{1953}{200}g_1^4\lambda_2 + \frac{117}{20}g_1^2g_2^2\lambda_2 + \frac{108}{5}g_1^2\lambda_2^2 + 108g_2^2\lambda_2^2 + \frac{9}{10}g_1^4\lambda_3 + \frac{15}{2}g_2^4\lambda_3 + \frac{12}{5}g_1^2\lambda_3^2\right.$$
$$+ 12g_2^2\lambda_3^2 + \frac{9}{20}g_1^4\lambda_4 + \frac{3}{2}g_1^2g_2^2\lambda_4 + \frac{15}{4}g_2^4\lambda_4 + \frac{12}{5}g_1^2\lambda_3\lambda_4 + \frac{6}{5}g_1^2\lambda_4^2 + 3g_2^2\lambda_4^2 + 12g_2^2\lambda_3\lambda_4 - 20\lambda_2\lambda_3\lambda_4$$
$$- 12\lambda_3^2\lambda_4 - \frac{3537}{2000}g_1^6 - \frac{1719}{400}g_1^4g_2^2 - \frac{303}{80}g_1^2g_2^4 - \frac{51}{8}g_2^4\lambda_2 - 312\lambda_2^3 - 12\lambda_2\lambda_4^2 - 16\lambda_3\lambda_4^2 - 20\lambda_2\lambda_3^2 - 8\lambda_3^3$$
$$\left. - 6\lambda_4^3 - \frac{12}{5}g_1^2\lambda_5^2 - 24\lambda_5^2Y_t^2 - 56\lambda_2\lambda_5^2 - 80\lambda_3\lambda_5^2 - 88\lambda_4\lambda_5^2 - 12\lambda_3^2Y_t^2 - 12\lambda_3\lambda_4Y_t^2 - 6\lambda_4^2Y_t^2\right) \tag{A17}$$

$$
\begin{aligned}
\beta_{\lambda_3}^{\mathrm{IDM}} = {}& \frac{1}{(4\pi)^2}\left(\frac{27}{100}g_1^4 + \frac{9}{4}g_2^4 + 12\lambda_1\lambda_3 + 12\lambda_2\lambda_3 + 4\lambda_3^2 + 4\lambda_1\lambda_4 + 4\lambda_2\lambda_4 + 2\lambda_4^2 + 8\lambda_5^2\right. \\
& \left. + 6\lambda_3 Y_t^2 - \frac{9}{10}g_1^2 g_2^2 - \frac{9}{5}g_1^2\lambda_3 - 9g_2^2\lambda_3\right) + \frac{1}{(4\pi)^2}\left(\frac{909}{200}g_1^4 g_2^2 + \frac{33}{40}g_1^2 g_2^4\right. \\
& + \frac{291}{8}g_2^6 + \frac{27}{10}g_1^4\lambda_1 + \frac{45}{2}g_2^4\lambda_1 + \frac{27}{10}g_1^4\lambda_2 + \frac{45}{2}g_2^4\lambda_2 + \frac{1773}{200}g_1^4\lambda_3 + \frac{33}{20}g_1^2 g_2^2\lambda_3 \\
& + \frac{72}{5}g_1^2\lambda_1\lambda_3 + 72g_2^2\lambda_1\lambda_3 + \frac{72}{5}g_1^2\lambda_2\lambda_3 + 72g_2^2\lambda_2\lambda_3 + \frac{6}{5}g_1^2\lambda_3^2 + 6g_2^2\lambda_3^2 + \frac{15}{2}g_2^4\lambda_4 \\
& + \frac{9}{10}g_1^4\lambda_4 + \frac{24}{5}g_1^2\lambda_1\lambda_4 + 6g_2^2\lambda_4^2 + 36g_2^2\lambda_1\lambda_4 + \frac{48}{5}g_1^2\lambda_5^2 + \frac{24}{5}g_1^2\lambda_2\lambda_4 + \frac{17}{4}g_1^2\lambda_3 Y_t^2 \\
& + \frac{45}{4}g_2^2\lambda_3 Y_t^2 + 40g_3^2\lambda_3 Y_t^2 + 36g_2^2\lambda_2\lambda_4 - 72\lambda_2\lambda_3^2 - 12\lambda_3^3 - \frac{3537}{1000}g_1^6 - 3g_1^2 g_2^2\lambda_1 \\
& - 3g_1^2 g_2^2\lambda_2 - 72\lambda_1\lambda_3^2 - \frac{111}{8}g_2^4\lambda_3 - 60\lambda_1^2\lambda_3 - 60\lambda_2^2\lambda_3 - \frac{9}{5}g_1^2 g_2^2\lambda_4 - 16\lambda_1^2\lambda_4 \\
& - 16\lambda_2^2\lambda_4 - 12g_2^2\lambda_3\lambda_4 - 32\lambda_1\lambda_3\lambda_4 - 32\lambda_2\lambda_3\lambda_4 - 4\lambda_3^2\lambda_4 - \frac{6}{5}g_1^2\lambda_4^2 - 28\lambda_1\lambda_4^2 \\
& - 28\lambda_2\lambda_4^2 - 16\lambda_3\lambda_4^2 - 12\lambda_4^3 - 144\lambda_1\lambda_5^2 - 144\lambda_2\lambda_5^2 - 72\lambda_3\lambda_5^2 - 176\lambda_4\lambda_5^2 \\
& \left. - \frac{171}{100}g_1^4 Y_t^2 - \frac{63}{10}g_1^2 g_2^2 Y_t^2 - \frac{9}{4}g_2^4 Y_t^2 - 72\lambda_1\lambda_3 Y_t^2 - 12\lambda_3^2 Y_t^2 - 24\lambda_1\lambda_4 Y_t^2 - 6\lambda_4^2 Y_t^2 - 24\lambda_5^2 Y_t^2 - \frac{27}{2}\lambda_3 Y_t^4\right)
\end{aligned}
\tag{A18}
$$

$$
\begin{aligned}
\beta_{\lambda_4}^{\mathrm{IDM}} = {}& \frac{1}{(4\pi)^2}\left(\frac{9}{5}g_1^2 g_2^2 + 4\lambda_1\lambda_4 + 4\lambda_2\lambda_4 + 8\lambda_3\lambda_4 + 4\lambda_4^2 + 32\lambda_5^2 + 6\lambda_4 Y_t^2 - \frac{9}{5}g_1^2\lambda_4 - 9g_2^2\lambda_4\right) \\
& + \frac{1}{(4\pi)^2}\left(6g_1^2 g_2^2\lambda_1 + 6g_1^2 g_2^2\lambda_2 + \frac{6}{5}g_1^2 g_2^2\lambda_3 + \frac{1413}{200}g_1^4\lambda_4 + \frac{153}{20}g_1^2 g_2^2\lambda_4\lambda_5^2 + \frac{63}{5}g_1^2 g_2^2 Y_t^2\right. \\
& + \frac{17}{4}g_1^2\lambda_4 Y_t^2 + \frac{45}{4}g_2^2\lambda_4 Y_t^2 + 40g_3^2\lambda_4 Y_t^2 + \frac{24}{5}g_1^2\lambda_1\lambda_4 + \frac{24}{5}g_1^2\lambda_2\lambda_4 \\
& + \frac{12}{5}g_1^2\lambda_3\lambda_4 + 36g_2^2\lambda_3\lambda_4 + \frac{24}{5}g_1^2\lambda_4^2 + 18g_2^2\lambda_4^2 + \frac{192}{5}g_1^2\lambda_5^2 + 216g_2^2\lambda_5^2 - 40\lambda_1\lambda_4^2 \\
& - 40\lambda_2\lambda_4^2 - 28\lambda_3\lambda_4^2 - \frac{657}{50}g_1^4 g_2^2 - \frac{42}{5}g_1^2 g_2^4 - \frac{231}{8}g_2^4\lambda_4 - 28\lambda_1^2\lambda_4 - 28\lambda_2^2\lambda_4 \\
& - 80\lambda_1\lambda_3\lambda_4 - 80\lambda_2\lambda_3\lambda_4 - 28\lambda_3^2\lambda_4 - 192\lambda_1\lambda_5^2 - 192\lambda_2\lambda_5^2 - 192\lambda_3\lambda_5^2 - 104\lambda_4 \\
& \left. - 24\lambda_1\lambda_4 Y_t^2 - 24\lambda_3\lambda_4 Y_t^2 - 12\lambda_4^2 Y_t^2 - 96\lambda_5^2 Y_t^2 - \frac{27}{2}\lambda_4 Y_t^4\right)
\end{aligned}
\tag{A19}
$$

$$
\begin{aligned}
\beta_{\lambda_5}^{\mathrm{IDM}} = {}& \frac{1}{(4\pi)^2}\left(4\lambda_1\lambda_5 + 4\lambda_2\lambda_5 + 8\lambda_3\lambda_5 + 12\lambda_4\lambda_5 + 6\lambda_5 Y_t^2 - \frac{9}{5}g_1^2\lambda_5 - 9g_2^2\lambda_5\right) \\
& + \frac{1}{(4\pi)^4}\left(\frac{1413}{200}g_1^4\lambda_5 + \frac{57}{20}g_1^2 g_2^2\lambda_5 + \frac{48}{5}g_1^2\lambda_3\lambda_5 + 36g_2^2\lambda_3\lambda_5 + \frac{72}{5}g_1^2\lambda_4\lambda_5\right. \\
& + 72g_2^2\lambda_4\lambda_5 + 24\lambda_5^3 + \frac{171}{200}g_1^4 Y_t^2 + \frac{9}{8}g_2^4 Y_t^2 + \frac{17}{4}g_1^2\lambda_5 Y_t^2 + \frac{45}{4}g_2^2\lambda_5 Y_t^2 \\
& + 40g_3^2\lambda_5 Y_t^2 + \frac{9}{2}\lambda_5 Y_t^4 - \frac{231}{8}g_2^4\lambda_5 - \frac{12}{5}g_1^2\lambda_1\lambda_5 - 28\lambda_1^2\lambda_5 - \frac{12}{5}g_1^2\lambda_2\lambda_5 \\
& - 28\lambda_2^2\lambda_5 - 80\lambda_1\lambda_3\lambda_5 - 80\lambda_2\lambda_3\lambda_5 - 28\lambda_3^2\lambda_5 - 88\lambda_1\lambda_4\lambda_5 - 88\lambda_2\lambda_4\lambda_5 \\
& \left. - 76\lambda_3\lambda_4\lambda_5 - 32\lambda_4^2\lambda_5 - \frac{63}{20}g_1^2 g_2^2 Y_t^2 - 24\lambda_1\lambda_5 Y_t^2 - 24\lambda_3\lambda_5 Y_t^2 - 36\lambda_4\lambda_5 Y_t^2\right)
\end{aligned}
\tag{A20}
$$

[1] G. Aad *et al.* (ATLAS Collaboration), Observation of a new particle in the search for the Standard Model Higgs boson with the ATLAS detector at the LHC, Phys. Lett. B **716**, 1 (2012).
[2] S. Chatrchyan *et al.* (CMS Collaboration), Observation of a new boson at a mass of 125 GeV with the CMS experiment at the LHC, Phys. Lett. B **716**, 30 (2012).
[3] G. Isidori, G. Ridolfi, and A. Strumia, On the metastability of the Standard Model vacuum, Nucl. Phys. **B609**, 387 (2001).
[4] F. Bezrukov, M. Y. Kalmykov, B. A. Kniehl, and M. Shaposhnikov, Higgs boson mass and new physics, J. High Energy Phys. 10 (2012) 140.
[5] G. Degrassi, S. Di Vita, J. Elias-Miro, J. R. Espinosa, G. F. Giudice, G. Isidori, and A. Strumia, Higgs mass and vacuum stability in the Standard Model at NNLO, J. High Energy Phys. 08 (2012) 098.
[6] D. Buttazzo, G. Degrassi, P. P. Giardino, G. F. Giudice, F. Sala, A. Salvio, and A. Strumia, Investigating the near-criticality of the Higgs boson, J. High Energy Phys. 12 (2013) 089.
[7] F. Csikor, Z. Fodor, and J. Heitger, End point of the hot electroweak phase transition, Phys. Rev. Lett. **82**, 21 (1999).
[8] K. Rummukainen, M. Tsypin, K. Kajantie, M. Laine, and M. E. Shaposhnikov, The universality class of the electroweak theory, Nucl. Phys. **B532**, 283 (1998).
[9] K. Kajantie, M. Laine, K. Rummukainen, and M. E. Shaposhnikov, Is there a hot electroweak phase transition at $m_H \gtrsim m_W$?, Phys. Rev. Lett. **77**, 2887 (1996).
[10] M. Gonderinger, Y. Li, H. Patel, and M. J. Ramsey-Musolf, Vacuum stability, perturbativity, and scalar singlet dark matter, J. High Energy Phys. 01 (2010) 053.
[11] M. Gonderinger, H. Lim, and M. J. Ramsey-Musolf, Complex scalar singlet dark matter: Vacuum stability and phenomenology, Phys. Rev. D **86**, 043511 (2012).
[12] O. Lebedev, On stability of the electroweak vacuum and the Higgs portal, Eur. Phys. J. C **72**, 2058 (2012).
[13] J. Elias-Miro, J. R. Espinosa, G. F. Giudice, H. M. Lee, and A. Strumia, Stabilization of the electroweak vacuum by a scalar threshold effect, J. High Energy Phys. 06 (2012) 031.
[14] P. Athron, J. M. Cornell, F. Kahlhoefer, J. Mckay, P. Scott, and S. Wild, Impact of vacuum stability, perturbativity and XENON1T on global fits of $\mathbb{Z}_2$ and $\mathbb{Z}_3$ scalar singlet dark matter, Eur. Phys. J. C **78**, 830 (2018).
[15] V. Barger, P. Langacker, M. McCaskey, M. Ramsey-Musolf, and G. Shaughnessy, Complex singlet extension of the Standard Model, Phys. Rev. D **79**, 015018 (2009).
[16] P. Borah and P. Ghosh, Unveiling the inert triplet desert region with a pNGB dark matter and its gravitational wave signatures, arXiv:2505.16521.
[17] M. Gonçalves, M. Mühlleitner, R. Santos, and T. Trindade, Dark matter in multi-singlet extensions of the Standard Model, arXiv:2505.07753.
[18] G. C. Branco, P. M. Ferreira, L. Lavoura, M. N. Rebelo, M. Sher, and J. P. Silva, Theory and phenomenology of two-Higgs-doublet models, Phys. Rep. **516**, 1 (2012).
[19] P. M. Ferreira, R. Santos, and A. Barroso, Stability of the tree-level vacuum in two Higgs doublet models against charge or *CP* spontaneous violation, Phys. Lett. B **603**, 219 (2004); **629**, 114(E) (2005).
[20] M. Maniatis, A. von Manteuffel, O. Nachtmann, and F. Nagel, Stability and symmetry breaking in the general two-Higgs-doublet model, Eur. Phys. J. C **48**, 805 (2006).
[21] S. Kanemura and Y. Mura, Criterion of perturbativity with the mass-dependent beta function in extended Higgs models, Phys. Rev. D **110**, 075016 (2024).
[22] B. Świeżewska, Yukawa independent constraints for two-Higgs-doublet models with a 125 GeV Higgs boson, Phys. Rev. D **88**, 055027 (2013); **88**, 119903(E) (2013).
[23] P. M. Ferreira and B. Swiezewska, One-loop contributions to neutral minima in the Inert Doublet Model, J. High Energy Phys. 04 (2016) 099.
[24] B. Swiezewska and M. Krawczyk, Diphoton rate in the Inert Doublet Model with a 125 GeV Higgs boson, Phys. Rev. D **88**, 035019 (2013).
[25] N. Chakrabarty, R. Roshan, and A. Sil, Two-component doublet-triplet scalar dark matter stabilizing the electroweak vacuum, Phys. Rev. D **105**, 115010 (2022).
[26] S. Melara-Duron, R. Gaitán, and J. M. Lamprea, Two dark matter candidates in a doublet-triplet Higgs model, Eur. Phys. J. Plus **140**, 154 (2025).
[27] M. Krawczyk, D. Sokolowska, P. Swaczyna, and B. Swiezewska, Constraining inert dark matter by $R_{\gamma\gamma}$ and WMAP data, J. High Energy Phys. 09 (2013) 055.
[28] L. L. Honorez, E. Nezri, J. F. Oliver, and M. H. G. Tytgat, The Inert Doublet Model: An archetype for dark matter, J. Cosmol. Astropart. Phys. 02 (2007) 028.
[29] N. Khan and S. Rakshit, Constraints on inert dark matter from the metastability of the electroweak vacuum, Phys. Rev. D **92**, 055006 (2015).
[30] X. J. Xu, Tree-level vacuum stability of two-Higgs-doublet models and new constraints on the scalar potential, Phys. Rev. D **95**, 115019 (2017).
[31] K. Kannike, Vacuum stability of a general scalar potential of a few fields, Eur. Phys. J. C **76**, 324 (2016); **78**, 355(E) (2018).
[32] S. Jangid and P. Bandyopadhyay, Distinguishing inert Higgs doublet and inert triplet scenarios, Eur. Phys. J. C **80**, 715 (2020).
[33] P. Bandyopadhyay, M. Frank, S. Parashar, and C. Sen, Interplay of inert doublet and vector-like lepton triplet with displaced vertices at the LHC/FCC and MATHUSLA, J. High Energy Phys. 03 (2024) 109.
[34] Nico Benincasa, Luigi Delle Rose, Kristen Kannike, and Luca Marzola, Multi-step phase transitions and gravitational waves in the Inert Doublet Model, J. Cosmol. Astropart. Phys. 12 (2022) 025.
[35] P. Bandyopadhyay and R. Mandal, Vacuum stability in an extended Standard Model with a leptoquark, Phys. Rev. D **95**, 035007 (2017).
[36] J. Braathen, M. Gabelmann, T. Robens, and P. Stylianou, Probing the Inert Doublet Model via vector-boson fusion at a muon collider, J. High Energy Phys. 05 (2025) 055.
[37] S. Banerjee, F. Boudjema, N. Chakrabarty, G. Chalons, and H. Sun, Relic density of dark matter in the Inert Doublet Model beyond leading order: The heavy mass case, Phys. Rev. D **100**, 095024 (2019).

[38] S. Yaser Ayazi and S. M. Firouzabadi, Footprint of triplet scalar dark matter in direct indirect search and invisible Higgs decay, Cogent Phys. **2**, 1047559 (2015).
[39] N. Khan, Exploring the hyperchargeless Higgs triplet model up to the Planck scale, Eur. Phys. J. C **78**, 341 (2018).
[40] P. Das, M. K. Das, and N. Khan, FIMP DM in the extended hyperchargeless Higgs triplet model, Phys. Rev. D **104**, 095026 (2021).
[41] P. Bandyopadhyay, S. Parashar, C. Sen, and J. Song, Probing Inert Triplet Model at a multi-TeV muon collider via vector boson fusion with forward muon tagging, J. High Energy Phys. 07 (2024) 253.
[42] P. Bandyopadhyay and A. Costantini, Obscure Higgs boson at colliders, Phys. Rev. D **103**, 015025 (2021).
[43] C. W. Chiang, G. Cottin, Y. Du, K. Fuyuto, and M. J. Ramsey-Musolf, Collider probes of real triplet scalar dark matter, J. High Energy Phys. 01 (2021) 198.
[44] N. Chakrabarty, D. K. Ghosh, B. Mukhopadhyaya, and I. Saha, Dark matter, neutrino masses, and high scale validity of an inert Higgs doublet model, Phys. Rev. D **92**, 015002 (2015).
[45] B. Swiezewska, Inert scalars and vacuum metastability around the electroweak scale, J. High Energy Phys. 07 (2015) 118.
[46] P. Bandyopadhyay, P. S. B. Dev, S. Jangid, and A. Kumar, Vacuum stability in inert Higgs doublet model with right-handed neutrinos, J. High Energy Phys. 08 (2020) 154.
[47] P. Bandyopadhyay and S. Jangid, Discerning singlet and triplet scalars at the electroweak phase transition and gravitational wave, Phys. Rev. D **107**, 055032 (2023).
[48] P. Bandyopadhyay and S. Jangid, Scrutinizing vacuum stability in IDM with type-III inverse Seesaw, J. High Energy Phys. 02 (2021) 075.
[49] D. Eriksson, J. Rathsman, and O. Stal, 2HDMC: Two-Higgs-doublet model calculator physics and manual, Comput. Phys. Commun. **181**, 189 (2010).
[50] C. Grojean, G. Servant, and J. D. Wells, First-order electroweak phase transition in the Standard Model with a low cutoff, Phys. Rev. D **71**, 036001 (2005).
[51] A. Ahriche, What is the criterion for a strong first-order electroweak phase transition in singlet models?, Phys. Rev. D **75**, 083522 (2007).
[52] J. R. Espinosa, T. Konstandin, and F. Riva, Strong electroweak phase transitions in the Standard Model with a singlet, Nucl. Phys. **B854**, 592 (2012).
[53] M. D. Astros, S. Fabian, and F. Goertz, Minimal inert doublet benchmark for dark matter and the baryon asymmetry, J. Cosmol. Astropart. Phys. 02 (2024) 052.
[54] S. AbdusSalam, M. J. Kazemi, and L. Kalhor, Upper limit on first-order electroweak phase transition strength, Int. J. Mod. Phys. A **36**, 2150024 (2021).
[55] D. Borah and J. M. Cline, Inert doublet dark matter with strong electroweak phase transition, Phys. Rev. D **86**, 055001 (2012).
[56] G. Gil, P. Chankowski, and M. Krawczyk, Inert dark matter and strong electroweak phase transition, Phys. Lett. B **717**, 396 (2012).
[57] J. M. Cline and K. Kainulainen, Improved electroweak phase transition with subdominant inert doublet dark matter, Phys. Rev. D **87**, 071701(R) (2013).
[58] S. S. AbdusSalam and T. A. Chowdhury, Scalar representations in the light of electroweak phase transition and cold dark matter phenomenology, J. Cosmol. Astropart. Phys. 05 (2014) 026.
[59] S. S. AbdusSalam and M. J. Kazemi, Electroweak phase transition in an inert complex triplet mode, Phys. Rev. D **103**, 075012 (2021).
[60] P. Bandyopadhyay, S. Jangid, and A. Karan, Constraining scalar doublet and triplet leptoquarks with vacuum stability and perturbativity, Eur. Phys. J. C **82**, 516 (2022).
[61] Y. Hamada, K. Kawana, and K. Tsumura, Landau pole in the Standard Model with weakly interacting scalar fields, Phys. Lett. B **747**, 238 (2015).
[62] A. D. Bond, G. Hiller, K. Kowalska, and D. F. Litim, Directions for model building from asymptotic safety, J. High Energy Phys. 08 (2017) 004.
[63] S. Heinemeyer, M. Mondragón, N. Tracas, and G. Zoupanos, Reduction of couplings and its application in particle physics, Phys. Rep. **814**, 1 (2019).
[64] M. A. May Pech, M. Mondragón, G. Patellis, and G. Zoupanos, Reduction of couplings in the Type-II 2HDM, Eur. J. Phys. **83**, 1129 (2023).
[65] R. Shrock, Study of the six-loop beta function of the $\lambda\varphi_4^4$ theory, Phys. Rev. D **94**, 125026 (2016).
[66] R. Shrock, Search for an ultraviolet zero in the seven-loop beta function of the $\lambda\phi_4^4$ theory, Phys. Rev. D **107**, 056018 (2023).
[67] M. Lüscher and P. Weisz, Scaling laws and triviality bounds in the lattice $\phi^4$ theory: (I). One-component model in the symmetric phase, Nucl. Phys. **B290**, 25 (1987).
[68] Oliver J. Rosten, Triviality from the exact renormalization group, J. High Energy Phys. 07 (2009) 019.
[69] M. Sher, Electroweak Higgs potential and vacuum stability., Phys. Rep. **179**, 273 (1989).
[70] A. M. Coutinho, A. Karan, V. Mirallesand A. Pich, Light scalars within the -conserving Aligned-two-Higgs-doublet model, J. High Energy Phys. 02 (2025) 057.
[71] H. Bahl, M. Carena, N. M. Coyle, A. Ireland, and C. E. M. Wagner, New tools for dissecting the general 2HDM, J. High Energy Phys. 03 (2023) 165.
[72] B. Q. Lu and D. Huang, Unitarity bounds on extensions of Higgs sector, J. High Energy Phys. 06 (2023) 209.
[73] Heather E. Logan, Lectures on perturbative unitarity and decoupling in Higgs physics, arXiv:2207.01064.
[74] I. F. Ginzburg and I. P. Ivanov, Tree-level unitarity constraints in the most general two Higgs doublet model, Phys. Rev. D **72**, 115010 (2005).
[75] M. V. Kompaniets and E. Panzer, Minimally subtracted six-loop renormalization of $O(n)$-symmetric $\phi^4$ theory and critical exponents, Phys. Rev. D **96**, 036016 (2017).
[76] Claudio Coriano, Luigi Delle Rose, and Carlo Marzo, Vacuum stability in $U(1)'$-extensions of the Standard Model with TeV scale right handed neutrinos, Phys. Lett. B **738**, 13 (2014).
[77] Claudio Coriano, Luigi Delle Rose, and Carlo Marzo, Constraints on Abelian extensions of the Standard Model

from two-loop vacuum stability and $U(1)_{B-L}$, J. High Energy Phys. 02 (2016) 135.
[78] F. Staub, SARAH 4: A tool for (not only SUSY) model builders, Comput. Phys. Commun. **185**, 1773 (2014).
[79] T. Katayose, S. Matsumoto, S. Shirai, and Y. Watanabe, Thermal real scalar triplet dark matter, J. High Energy Phys. 09 (2021) 044.
[80] G. Aad *et al.* (ATLAS Collaboration), Measurement of the Higgs boson mass with $H->\gamma\gamma$ decays in 140 fb$^{-1}$ of $\sqrt{s}=13$ TeV pp collisions with the ATLAS detector, Phys. Lett. B **847**, 138315 (2023).
[81] S. Navas *et al.* (Particle Data Group), Review of particle physics, Phys. Rev. D **110**, 030001 (2024).
[82] P. Bandyopadhyay and S. Parashar, Dark clouds to silver linings over the hyperchargeless scalar triplets, arXiv: 2505.23257.
[83] T. Biekötter, S. Heinemeyer, J. M. No, K. Radchenko, M. O. O. Romacho, and G. Weiglein, First shot of the smoking gun: Probing the electroweak phase transition in the 2HDM with novel searches for $A->ZH$ in $\ell^+\ell^-t\bar{t}$ and $\nu\nu b\bar{b}$, J. High Energy Phys. 01 (2024) 107.
[84] J. Aalbers *et al.* (LZ Collaboration), Dark matter search results from 4.2 tonne-years of exposure of the LUX-ZEPLIN (LZ) experiment, arXiv:2410.17036.
[85] A. Datta, N. Ganguly, N. Khan, and S. Rakshit, Exploring collider signatures of the inert Higgs doublet model, Phys. Rev. D **95**, 015017 (2017).
[86] P. Bandyopadhyay, E. J. Chun, and R. Mandal, Scalar dark matter in leptophilic two-Higgs-doublet model, Phys. Lett. B **779**, 201 (2018).
[87] P. Bandyopadhyay, E. J. Chun, and R. Mandal, Phenomenology of Higgs bosons in inverse seesaw model with type-X two Higgs doublet at the LHC, J. High Energy Phys. 08 (2019) 169.
[88] P. Bandyopadhyay, K. Huitu, and A. Sabanci Keceli, Multi-lepton signatures of the triplet like charged Higgs at the LHC, J. High Energy Phys. 05 (2015) 026.
[89] P. Bandyopadhyay, C. Coriano, and A. Costantini, Probing the hidden Higgs bosons of the $Y=0$ triplet- and singlet-extended supersymmetric Standard Model at the LHC, J. High Energy Phys. 12 (2015) 127.
[90] P. Bandyopadhyay, C. Corianò, and A. Costantini, General analysis of the charged Higgs sector of the $Y=0$ triplet-singlet extension of the MSSM at the LHC, Phys. Rev. D **94**, 055030 (2016).
[91] P. J. Fox, R. Harnik, J. Kopp, and Y. Tsai, Missing energy signatures of dark matter at the LHC, Phys. Rev. D **85**, 056011 (2012).
[92] G. G. da Silveira and M. S. Mateus, Investigation of scalar and fermion dark matter in mono-photon production at high-energy colliders, Eur. Phys. J. C **84**, 181 (2024).
[93] S. Schramm (ATLAS Collaboration), ATLAS sensitivity to WIMP dark matter in the monojet topology at $\sqrt{s}=$ TeV, Nucl. Part. Phys. Proc. **273–275**, 2397 (2016).
[94] S. Chatrchyan *et al.* (CMS Collaboration), Search for new physics with a monojet and missing transverse energy in $pp$ collisions at $\sqrt{s}=7$ Tev, Phys. Rev. Lett. **107**, 201804 (2011).
[95] P. Bandyopadhyay and S. Parashar, Probing a scalar singlet-triplet extension of the Standard Model via vector boson fusion at a muon collider, Phys. Rev. D **110**, 115032 (2024).
[96] A. Hayrapetyan *et al.* (CMS Collaboration), Search for exotic decays of the Higgs boson to a pair of pseudoscalars in the $\mu\mu$bb and $\tau\tau$bb final states, Eur. Phys. J. C **84**, 493 (2024).
[97] V. Khachatryan *et al.* (CMS Collaboration), Search for resonant pair production of Higgs bosons decaying to two bottom quark–antiquark pairs in proton–proton collisions at 8 TeV, Phys. Lett. B **749**, 560 (2015).
[98] A. Hayrapetyan *et al.* (CMS Collaboration), Constraints on the Higgs boson self-coupling from the combination of single and double Higgs boson production in proton-proton collisions at $s=13$ TeV, Phys. Lett. B **861**, 139210 (2025).
[99] G. Aad *et al.* (ATLAS Collaboration), Search for pair production of boosted Higgs bosons via vector-boson fusion in the $b\bar{b}b\bar{b}$ final state using $pp$ collisions at $\sqrt{s}$-13 TeV with the ATLAS detector, Phys. Lett. B **858**, 139007 (2024).
[100] G. Aad *et al.* (ATLAS Collaboration), Search for triple Higgs boson production in the 6b final state using pp collisions at $s=13$ TeV with the ATLAS detector, Phys. Rev. D **111**, 032006 (2025).
[101] G. Aad *et al.* (ATLAS Collaboration), Search for a resonance decaying into a scalar particle and a Higgs boson in the final state with two bottom quarks and two photons in proton–proton collisions at $\sqrt{s}=$ 13 TeV with the ATLAS detector, J. High Energy Phys. 11 (2024) 047.
[102] G. Aad *et al.* (ATLAS Collaboration), Combination of searches for resonant Higgs boson pair production using pp collisions at $s=13$ TeV with the ATLAS detector, Phys. Rev. Lett. **132**, 231801 (2024).